\documentclass{aa}     
\usepackage{graphicx}

\newcommand{\Mpc}{$h^{-1}$\thinspace Mpc}

\begin{document}

\title{Environmental Enhancement of Loose Groups 
around Rich Clusters of Galaxies} 

\author {M. Einasto\inst{1}, J. Einasto\inst{1}, V. M\"uller\inst{2},
 P. Hein\"am\"aki\inst{1,3},  \& D.L. Tucker\inst{4}} 
\authorrunning{M. Einasto et al.}
\offprints{M. Einasto }
\institute{ Tartu Observatory, EE-61602 T\~oravere, Estonia
\and
Astrophysical Institute Potsdam, An der Sternwarte 16,
D-14482 Potsdam, Germany
\and
Tuorla Observatory, V\"ais\"al\"antie 20, Piikki\"o, Finland 
\and
Fermi National Accelerator Laboratory, MS 127, PO Box 500, Batavia, 
IL 60510, USA
}

\date{ Received   2002; Accepted ...  } 
\titlerunning{Environmental enhancement of loose groups}
\authorrunning{Einasto et al.}

\abstract{ We have studied the properties of Las Campanas Loose Groups
(Tucker et al. \cite{tuc:tuc}) 
in the neighbourhood of rich (Abell, APM and
X-ray) clusters of galaxies.  These loose groups show strong
evidence of segregation measured in terms of the group richness and
the group velocity dispersion: loose groups in the neighbourhood of
a rich cluster are typically $2.5$ times more massive and $1.6$ times
more luminous than groups on average, and these loose groups have
velocity dispersions $1.3$ times larger than groups on average.  
This is evidence that the large-scale gravitational field causing 
the formation of rich clusters enhances the
evolution of neighbouring poor systems, a phenomenon recently
established in numerical simulations of group and cluster formation.
\keywords{cosmology: observations -- cosmology: large-scale structure
of the Universe} }

\maketitle

\section{Introduction}

On large scales galaxies and clusters form filamentary superclusters,
leaving the space between these filamentary structures almost devoid
of galaxies.  Superclusters and voids form an irregular pattern, which
we call the supercluster-void network (Einasto et al. \cite{e2001} and
references therein). The fine structure of superclusters, as well as
the distribution of matter in low density regions between
superclusters, can be studied using catalogues of galaxies and of poor
systems of galaxies (Lindner et al. \cite{lind95}). Early wide-field
catalogues of galaxies and groups, however, contained only nearby
objects [see references in Tucker et al. (\cite{tuc:tuc}; hereafter
TUC)].  This has changed recently:
several deep surveys of galaxies are now  available and can be
used for studies of the large scale structure of the Universe in
greater detail and on larger scales than ever before.  Among these
surveys are the ESO Key Program Survey (Vettolani et
al. \cite{vet97}), the Las Campanas Redshift Survey (LCRS; Shectman et
al. \cite{she:she}), the 2 degree Field Galaxy Redshift Survey (2dFGRS)
(Colless et al. \cite{col01}) and the Sloan Digital Sky Survey (York
et al. \cite{yo0}). On the basis of these and other surveys
several deep catalogues
of groups have been compiled recently: a catalogue
of groups from the ESO Key Program Survey (Ramella et al. \cite{ram99}),
CNOC2 groups (Carlberg et al. \cite{car01}), and groups 
from the publicly available data of the
2dFGRS survey (Merchan \& Zandivarez \cite{mer02}).

In this paper, we investigate the environments surrounding the Las
Campanas Loose Groups (LCLGs), a catalogue of 1495 loose groups
extracted from the LCRS survey (TUC). This catalogue provides  an
opportunity to study the space distribution and the properties of
loose groups on large scales, out to distances of $450$~\Mpc.

Rich clusters of galaxies represent high density enhancements in the
distribution of galaxies. Therefore, using samples of rich clusters
and of neighbouring loose groups, we can study the possible influence
of high density environments on the properties of loose groups.  In
order to pursue this study, we have investigated properties of LCLGs
in the vicinity of rich clusters of galaxies, including clusters from
the Abell and APM catalogues, X-ray clusters, and even a set of the
richest groups from the LCLG catalogue itself.  We have then compared
the properties of these dense-environment LCLGs with those of typical
LCLGs.\footnote{The three-dimensional distribution of LCLGs and rich
clusters can be seen by visiting the home page of the Tartu
Observatory ({\tt http://www.aai.ee/$\sim$maret/cosmoweb.html}).}

In the next section we describe the samples used.  In Sect. 3 we
extract samples of loose groups near rich clusters, and in Sect. 4 we
will study the properties of these loose groups.  In the last 
section we will discuss and summarise of our results.

\section{Observational data}

\subsection{LCLGs}

The LCRS (Shectman et al. \cite{she:she} ) is an optically selected
galaxy redshift survey that extends to a redshift of 0.2 and includes
6 slices, each covering an area of roughly $1.5 \times 80$ degrees.
Three of these slices are located in the Northern Galactic Cap and are
centred at declinations $\delta=-3^{\circ}, -6^{\circ}, -12^{\circ}$;
the other three slices are located in the Southern Galactic Cap and are
centred at declinations $\delta= -39^{\circ}, -42^{\circ},
-45^{\circ}$.  The thickness of the slices is approximately $7.5$~\Mpc\ at the
survey's median redshift.  In all, the LCRS contains $23,697$
galaxies with redshifts within its official photometric and geometric
boundaries.

Survey spectroscopy was first carried out via a 50 fibre multiobject
spectrograph,  and the nominal apparent magnitude limits for
the spectroscopic fields were $16.0 \le R \le 17.3$.  Partway
through the survey, the spectrograph was upgraded to 112 fibres, which
permitted a selection of galaxies over the somewhat larger range of
apparent magnitudes of $15.0 \le R \le 17.7$.  For the sake of
efficiency, each LCRS spectroscopic field was observed only once;
fields which were observed with the 50-fibre spectrograph were not
re-observed.  Therefore, the selection criteria varied from field to
field, often within a given slice.

Using a friends-of-friends percolation algorithm, TUC extracted the
LCLG catalogue from the LCRS.  The linking length parameters were chosen
so that each group is contained within a galaxy number density
enhancement contour of $\delta n/n = 80$.  In extracting these LCLGs,
great care was taken in order to account for both the radial selection
function and the field-to-field selection effects inherent in the
LCRS.  This care is evident in that the derived properties of the
LCLGs in the 50-fibre fields do not differ substantially from the
derived properties of the LCLGs in the 112-fibre fields (see TUC for
details).

The LCLG catalogue contains 1495 groups in a redshift range of
$10,000 \le cz \le 45,000$~{\rm km s}$^{-1}$.  This is one of the
first deep, wide samples of loose groups; as such, it enables us for
the first time to investigate the space distribution and properties of
groups in a large volume.

\subsection{Abell-class groups from the LCLG catalogue}

We chose the first sample of rich clusters from the LCLG catalogue itself.
TUC have calculated an estimate of the 
Abell counts $N_{\rm ACO}$ for each group.  
Among these groups we choose those loose groups 
with $N_{\rm ACO} \geq 30$, which corresponds to an Abell
richness class of $R = 0$.  

However, this sample may be affected by selection effects. The study 
of the mass function of LCLGs (Hein\"am\"aki et al. \cite{hei:hei}) 
shows that the sample of loose groups from TUC is complete in the case 
of groups with  masses exceeding $10^{13.5} - 10^{14} M_{\sun}$. 
Thus in order to obtain a complete sample of  Abell-class groups from 
the LCLG list, we excluded from the sample of  Abell-class groups all 
groups with masses less than $10^{13.8} M_{\sun}$, leaving  
56  Abell-class LCLGs.

The advantage in using this sample is that these clusters have been
determined in the same way as the other groups in the catalogue.
Thus, their measured properties can be easily compared with those of
other LCLGs without the need to contend with various unknown
inter-catalogue systematics. Furthermore, all the neighbours of 
Abell-class loose groups from the LCLG list are true neighbours, and not
mere positional coincidences between rich clusters and loose groups.

\subsection{Abell clusters and superclusters}

We use Abell's catalogue of rich clusters (Abell \cite{abell} and 
Abell et al. \cite{aco}), exploiting Andernach \& Tago's (\cite{at98}) 
recent compilation of all published redshifts for Abell cluster 
galaxies.  This compilation contains all known Abell clusters with 
measured redshifts, based on redshifts of individual cluster galaxies, 
and redshift estimates of  clusters according to the formula derived 
by Peacock \& West (\cite{pea92}), for both Abell catalogues 
(Abell \cite{abell} and Abell et al. \cite{aco}).  
We omitted from the compilation all supplementary, or 
S-clusters, but included clusters of richness class $R=0$ from the 
main catalogue.  From the general Abell cluster sample we selected all 
clusters with measured redshifts up to $z_{\rm lim}=0.13$; beyond 
this limit the fraction of clusters with measured redshifts becomes 
small. Our sample contains 1663 clusters, 1071 of which have measured 
redshifts.  We consider that a cluster has a measured redshift if at 
least one of its member galaxy has a measured redshift.  In cases 
where the cluster has less than three galaxies with measured 
redshifts, and the measured and estimated redshifts differ by more 
than a factor of two ($|\log(z_{\rm meas}/z_{\rm est})| > 0.3$), the estimated 
redshift was used. In 
this compilation the redshifts of Abell clusters in the region of the LCRS 
were corrected taking into account the redshift data from the LCRS itself. As a 
result, in the present study only one Abell cluster has no 
measured redshift, and this cluster (Abell 2031) is located outside 
the LCRS borders.

Note that some Abell clusters matched to LCLGs by TUC are absent from 
our list.  TUC used an angular separation criterion to match Abell 
clusters with LCLGs ($\Delta\theta < 12$~arc-min), whereas physical 
separation criteria are used here ($\Delta s <$ 6~\Mpc, as described
in the next section). 
However, the main reason for the absence of some clusters is 
that loose groups were determined in a narrower redshift range than 
the whole survey ($10,000 \le cz \le 45,000$~{\rm km s}$^{-1}$,
while the whole survey extends up to redshifts of about
$60,000$~{\rm km s}$^{-1}$).  
Nearby and distant clusters that fall within the survey boundaries were 
of course not included  in the present study. Note also that some of 
the rich clusters  that have nearby LCLGs are 
actually located outside of the borders of LCRS slices. 
Altogether there are 64 Abell clusters used in the present study, 
34 of them are of richness class $R = 0$.

The sample of Abell clusters is described in more
detail by Einasto et al. (\cite{e2001}), where we present an updated
catalogue of superclusters composed of Abell clusters.

\subsection{X-ray selected cluster samples}

We also use X-ray clusters found in the ROSAT All-sky Survey (RASS,
Tr\"umper \cite{tru:tru}).  On the basis of RASS, several catalogues
of X-ray selected galaxy clusters have been prepared.  In the present
paper we shall use these samples of X-ray clusters for which the
sample volumes are intersected by LCRS slices:

\begin{enumerate}

\item Clusters from the all-sky ROSAT Bright Survey of high galactic
latitude RASS sources. A detailed description of the data is given in
Voges et al. \cite{vog:vog}, and the catalogue of X-ray clusters, AGNs,
galaxies, small groups of galaxies and other objects is described in
Schwope et al. (\cite{sch:sch}).  We shall refer to this sample as the 
RBS sample.

\item A flux-limited sample of bright clusters from the Southern sky 
(de Grandi et al. \cite{deg:deg}; see also Guzzo et al. 
\cite{guz:guz} and Borgani \& Guzzo \cite{bog:bog}).

\end{enumerate}

All 14 X-ray clusters in the present study  have measured redshifts.
For details we refer to Einasto et al. (\cite{e2001}).

\subsection{APM clusters}

The APM cluster catalogue (Dalton et al. \cite{dal:dal}; hereafter D97)
was derived from the APM galaxy catalogue (Maddox et al. 
\cite{mes:mes}), which itself was extracted from plates
scanned by the Automatic Plate Measuring (APM) Facility.

The D97 APM cluster catalogue contains 957 clusters, of which 374 have
measured redshifts.  The redshifts of APM clusters in the region of
LCRS were corrected taking into account the redshift data from LCRS.
In Einasto et al. (\cite{e2002}) we analysed selection effects in the
APM catalogue and found that in the case of APM clusters with
estimated redshifts, their space density is artificially enhanced at
distances of about $300 - 350$~\Mpc\ due to selection effects. This
property may affect our present analysis. Therefore we decided to use
in the present study only those APM clusters with measured velocities.

The APM cluster catalogue contains clusters which are typically poorer
than clusters in the Abell catalogue, and have
a higher space density. The APM clusters
with measured redshifts are less affected by projection effects 
than Abell clusters (see Einasto et al. \cite{e2002}).  
In what follows below, 55 APM clusters  
provide an additional hunting ground for LCLGs.

For consistency with TUC and our earlier studies of superclusters we
calculated distances to groups and clusters using the following
formula (Mattig \cite{mat58}):
$$
r = {c \over {H_0 q_02}} {{q_0 z + (q_0-1)(\sqrt{1+2q_0z} -1)}
\over {1+z}};
\eqno(1)
$$
where $c$ is the velocity of light, $H_0$ is the Hubble constant, and
$q_0$ is the deceleration parameter.  As in TUC we use
$H_0 = 100~h$~km~s$^{-1}$~Mpc$^{-1}$, and $q_0=0.5$. Modern data
suggest a model with cosmological constant, which gives about 10~\%
larger distances than the high-density model used previously.

\section{Populations of LCLGs around rich clusters}

To extract samples of loose groups near rich clusters
we searched for LCLGs in spheres around rich clusters using a wide
range of neighbourhood radii. 

The lower limit of the search radius is determined by the virial radii 
of the rich clusters and the harmonic radii of the loose groups. For most 
rich clusters in our samples virial radii are not available. It is 
possible to estimate virial radii of clusters using the data about 
their velocity dispersions (Carlberg et al. 1996) or richness (Mazure 
et al. 1996). However, the scatter of such estimates is very high;  
moreover, these estimates are only (relatively) reliable for clusters 
of richness class $R \geq 1$, whereas about half 
of the Abell  clusters in our samples 
are of richness class $R = 0$.

\begin{figure}
\vspace{7cm}
\caption{The estimate of virial radii of Abell clusters (in \Mpc),
against the richness values of these clusters. Open circles:
clusters with virial radii from literature (see text). 
Dots: estimate of virial radii for Abell clusters
in our sample. }
\includegraphics{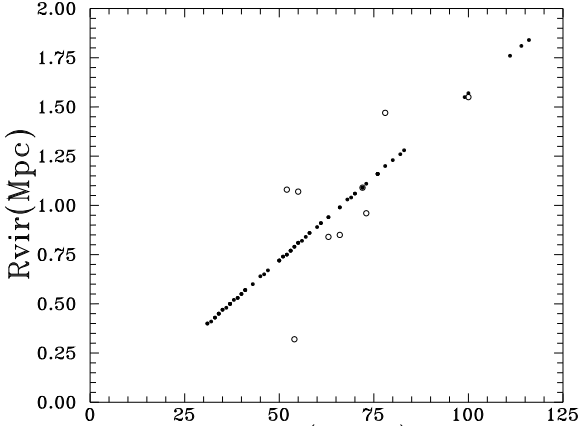}
\label{fig:ngalrv}
\end{figure}

Thus, we used another approach. We applied virial radii for some Abell
clusters in our sample determined by Girardi et
al. (\cite{gir98}). 
These radii are plotted against cluster richness values in
Fig.~\ref{fig:ngalrv}. We estimated virial radii for other clusters in
our sample using these data and cluster's richness values, keeping in mind
that we only need estimates of these radii in order to estimate
the lower limit of search radius for loose groups near rich clusters.
Fig.~\ref{fig:ngalrv} shows that neighbours closer than
approximately $1.6$~\Mpc\ should not be included into our sample.
The three richest Abell clusters have larger virial radii estimates, but
these clusters do not have very close neighbours among loose groups
(the closest neighbour group is located at a distance of $2.7$~\Mpc\
from the centre of one of these clusters).  We used this limit,
$1.6$~\Mpc, in the case of all samples of rich clusters. 
Additionally, we checked the harmonic radii of loose groups.  The loose
groups included into our sample all have harmonic radii less than half
the distance between a rich cluster and the loose group.

An upper limit of the search radius is determined by the large
scale distribution of LCRS galaxies and loose groups in the region of
rich clusters. When we increase the search radius, then at certain
radii we begin to include into our sample loose groups that are
separated from rich clusters by voids. Thus if we simply increase the
search radius we obtain populations of groups that actually may not be
related to rich cluster. If we want to study the loose groups in the
neighbourhood of rich clusters at large radii, we should analyse the
region of each cluster individually, and take into account the large
scale distribution of galaxies and groups.  For the present study we
found that an appropriate upper limit for the radius of the neighbour
search is $6$\Mpc.  By using a search radius of 6~\Mpc\ we obtain a
population of loose groups that surround rich clusters and lie within
the same high density regions (i.e., within the same
superclusters). In some cases, rich clusters within a search radius of
$6$ \Mpc\ are actually located outside of the borders of LCRS slices.

Additionally, we checked for possible line-of-sight alignments between
loose groups and rich clusters using the following test.  For a given
rich cluster, we searched within a narrow cone 
for any neighbors within a $\pm 6$\Mpc\ line-of-sight
distance from the cluster's location.  If there were neighbours in
this cone, then in order to be included into the final list of
neighbours the distance between the rich cluster and a loose group had
to be at least 3.2~\Mpc\ (i.e., twice the original search radius limit
described above).  With this procedure, the largest number of
neighbours excluded from the initial sample -- 20 -- was among the
neighbours of APM clusters. According to the loose group richnesses
and velocity dispersions, fifteen of these loose groups were possible
matches with the APM clusters.  In the case of Abell clusters there
were six neighbours excluded in this way, which were probable matches
between Abell clusters and loose groups (taking into account the
parameters of those groups). Around X-ray clusters, as a result of
this test we excluded three neighbours from the final sample.  And
finally, we excluded seven neighbours of  Abell-class loose groups located
along a line-of-sight with these groups.  This procedure excluded all the
groups within $2$~\Mpc\ of the centre of a rich cluster.

In this way we have defined our samples of loose groups which
neighbour rich clusters.  To aid our analysis, we have also defined
for each of these test samples a corresponding comparison sample
consisting of all those loose groups which do {\em not} neighbour a
rich cluster.  We excluded from these comparison samples all 
loose groups that are within $\pm 6$\Mpc\ of a rich cluster.

\begin{table}
\begin{center}
\caption{The list of LCLG systems around  APM clusters}  
\tabcolsep 2pt 
\tiny
\begin{tabular}{rrrrrr}
\hline
$N_{APM}$&$\alpha$ & $\delta$ & $D$ & $N_{LCLG}$  &Near \\
(1)& (2) & (3) & (4) & (5)&(6)   \\
\hline
\multispan3 Slice $\delta=-39^{\circ}$   &&&\\
\hline
  53  & 00 20  & -38 21  & 328.1  &   8, 9      &  N  \\
 160  & 01 15  & -38 13  & 218.6  &  39         &  N    \\
 162  & 01 15  & -36 51  & 213.2  &  37         &  N    \\
 172  & 01 23  & -39 43  & 242.6  &  44         &       \\
 173  & 01 23  & -38 11  & 223.9  &  46, 49     & \\ 
 413  & 03 32  & -39 15  & 172.4  & 116, 117, 122    & N \\
 415  & 03 32  & -38 57  & 177.9  & 111, 114, 122, 123    &   \\
 414  & 03 32  & -39 34  & 287.0  &  115        &        \\
 654  & 21 15  & -39 47  & 163.9  &  169        &  N  \\
 677  & 21 30  & -38 47  & 365.4  &   171, 174  & N \\
 800  & 22 34  & -39 23  & 207.6  &   198       & N \\
 873  & 23 13  & -39 13  & 186.2  &   223, 228  &  N \\
 914  & 23 35  & -38 30  & 300.0  &   239       &  N \\
 944  & 23 53  & -39 42  & 284.5  &   249, 251, 256 &  N \\
 947  & 23 57  & -39 46  & 281.9  &   249, 256  & N \\
\hline
\multispan3 Slice $\delta=-42^{\circ}$   &&&\\
\hline
  44  & 00 16  & -42 03  & 258.4  &    16      &        \\
 139  & 01 01  & -43 07  & 153.1  &   41, 42, 44 &        \\
 184  & 01 29  & -41 12  & 245.2  &   51         &  N   \\
 189  & 01 30  & -42 26  & 242.6  &    50        &  N     \\
 360  & 03 14  & -42 55  & 183.4  &    92, 95    &  N     \\
 365  & 03 15  & -42 16  & 177.9  &    92, 100   & N \\
 387  & 03 20  & -41 30  & 183.4  &    95        &   \\
 443  & 03 43  & -41 21  & 169.6  &   105, 106, 110 &  \\
 630  & 21 03  & -42 43  & 300.0  &   140, 141   &     \\   
 644  & 21 12  & -42 50  & 215.9  &   148, 150   &      \\
 673  & 21 28  & -43 30  & 292.3  &   156        &       \\
 688  & 21 34  & -41 19  & 185.9  &   165        &   \\
 878  & 23 15  & -42 38  & 315.4  &   234        &  \\    
 877  & 23 14  & -42 57  & 268.9  &   238        &       \\
 956  & 23 59  & -44 07  & 116.7  &   261, 5     &     N    \\
\hline
\multispan3 Slice $\delta=-45^{\circ}$   &&&\\
\hline
  19  & 00 11  & -45 21  & 375.6  &    10          &  N  \\
 139  & 01 02  & -43 08  & 153.1  &    32          &  N  \\
 185  & 01 28  & -44 32  & 345.7  &    57          &     \\
 238  & 02 06  & -45 01  & 289.6  &    75, 78, 81  &     \\
 289  & 02 43  & -45 26  & 271.5  &    95,  101    &     \\
 351  & 03 12  & -44 47  & 315.3  &   111          &  N  \\      
 362  & 03 14  & -45 40  & 202.4  &   112          &  N  \\
 366  & 03 15  & -45 14  & 213.2  &   114          &     \\
 369  & 03 16  & -44 50  & 215.9  &   114          &  N  \\
 374  & 03 16  & -44 25  & 205.1  &   112, 113     &  N  \\
 433  & 03 39  & -45 51  & 188.8  &   115, 119, 120, 124, 125 &     \\
 450  & 03 46  & -45 40  & 197.0  &   124, 129, 130 &   \\
 509  & 04 30  & -46 13  & 191.5  &   151           &  N  \\
 642  & 21 10  & -44 46  & 284.5  &   165           &     \\
 650  & 21 14  & -45 09  & 261.1  &   167           &  N  \\
 651  & 21 14  & -45 41  & 188.9  &   159, 173      &     \\
 653  & 21 15  & -45 28  & 271.5  &   169           &     \\
 657  & 21 16  & -45 32  & 276.8  &   169           &     \\
 659  & 21 18  & -45 43  & 271.5  &   169           &     \\
 709  & 21 44  & -44 07  & 178.0  &   184           &  N  \\
 757  & 22 09  & -45 44  & 328.1  &   198           &  N  \\
 812  & 22 41  & -45 35  & 253.2  &   217, 219, 221, 223    &  N  \\
 814  & 22 42  & -45 21  & 253.2  &   217, 219, 221, 223    &     \\
 825  & 22 47  & -45 35  & 147.6  &   228, 229              &  N  \\
 844  & 22 58  & -44 16  & 242.6  &   230,  235       &     \\
 915  & 23 37  & -46 15  & 191.6  &   250, 254       &  N  \\
 956  & 23 59  & -44 07  & 116.7  &   263 &             \\
\hline
\label{tab:lgapm}
\end{tabular}
\end{center}
 
{\it Columns are given in Table~\ref{tab:lga}}
                            
\end{table}

\begin{table}
\begin{center}
\caption{The list of LCLG systems around  X-ray clusters}
\tabcolsep 2pt 
\tiny
\begin{tabular}{rrrrrr}
\hline
$N_{RBS}$&$\alpha$ & $\delta$ & $D$ & $N_{LCLG}$ & $N_{Abell}$ \\
(1)& (2) & (3) & (4) & (5) &(6)    \\
\hline
\multispan3 Slice $\delta=-3^{\circ}$  &&&\\
\hline
 1193  &  12 56 &  -1 29 & 238.3  & 174 &            A1650      \\
 1197  &  12 56 &  -3 55 & 236.7  & 172 &            A1651      \\
 1205  &  13 00 &  -2 15 & 231.9  & 172, 174, 176  & A1663      \\
\hline  
\multispan3 Slice $\delta=-12^{\circ}$   &&&\\
\hline
  851  & 10 15 &  -10 27 & 168.6  &  7, 8    &      A 970 \\
 1020  & 11 38 &  -12 05 & 323.5  &  94 &           A1348     \\
 1151  & 12 42 &  -11 43 & 264.9  & 143, 144, 147    & A1606 \\
 1337  & 13 59 &  -10 55 & 198.1  & 218    &           A1837   \\
\hline  
\multispan3 Slice $\delta=-39^{\circ}$   &&&\\
\hline
  521  & 04 14 &  -38 06 & 143.9  &  141, 142, 143, 147    &  nA \\
\hline  
\multispan3 Slice $\delta=-42^{\circ}$   &&&\\
\hline
  469  & 03 45 &  -41 12 & 172.2  & 105, 106, 110      &  A0384   \\
 1990  & 23 18 &  -42 10 & 252.1  &  236, 237          &  A3998    \\
\hline  
\multispan3 Slice $\delta=-45^{\circ}$   &&&\\
\hline
  459  & 03 40   &  -45 41 & 191.3  & 115, 120, 127   &nA \\
 1782  & 21 43   &  -44 08 & 177.7  & 187 &          A3809 \\
 Rx31  & 03 12   &  -45 36 & 206.7  & 112 &          A3104 \\
\hline
\label{tab:lgx}
\end{tabular}
\end{center}

{\it Columns are given in Table~\ref{tab:lga}}
\end{table}

We denote samples of LCLGs as follows:
\begin{itemize}  

\item The sample of groups around Abell-class LCLGs is denoted by LCLG.LG 
      (the corresponding comparison sample by LCLG.cmp.LG).  

\item The sample of groups around Abell clusters is denoted by LCLG.Abell 
      (the corresponding comparison sample by LCLG.cmp.Abell).  

\item The sample of groups around APM clusters is denoted by LCLG.APM 
      (the corresponding comparison sample by LCLG.cmp.APM).  

\item The sample of groups around X-ray clusters is denoted by LCLG.X
      (the corresponding comparison sample by LCLG.cmp.X)
.
\end{itemize}

The lists of LCLGs around Abell, APM and X-ray clusters are given in 
Table~\ref{tab:lga}, Table~\ref{tab:lgapm} and Table~\ref{tab:lgx}.

\begin{table*}
\caption{Median and upper quartile (in parentheses) values of LCLG properties.}

\begin{tabular}{lrrrrrrr}
Sample & $N_{\rm group}$ & $N_{\rm obs}$ & $N_{\rm ACO}$ & $R_{\rm h}$
&  $\sigma_{\rm los}$ 
& $\log M_{\rm vir}$ & $\log L_{\rm tot}$  \\
& & & &\Mpc\ 
&   km~s$^{-1}$ & $h^{-1}M_{\odot}$  & $h^{-2}L_{\odot}$\\
(1)& (2) & (3) & (4) & (5)& (6) &(7) & (8)  \\

\hline 
LCLG.LG         & 95 & 4.5 (7.0)& 19.5 (28.0)& 0.65 (0.89)& 235 (295)& 13.65 (14.00)&11.35 (11.65) \\
LCLG.cmp.LG   & 1340 & 4.0 (5.5)& 15.5 (25.5)& 0.46 (0.75)& 175 (275)& 13.20 (13.75)&11.15 (11.35) \\
&&&&&&& \\
LCLG.Abell      & 96 & 5.5( 9.5)& 18.5 (30.0)& 0.51 (0.78)& 215 (300)& 13.55 (13.90)&11.25 (11.60) \\
LCLG.cmp.Abell& 1379 & 4.5 (5.5)& 16.5 (24.5)& 0.46 (0.76)& 180 (265)& 13.25 (13.75)&11.15 (11.45) \\
&&&&&&& \\
LCLG.X          & 24 & 6.0(10.0)& 15.0 (30.0)& 0.54 (0.80)& 225 (400)& 13.65 (14.15)&11.15 (11.75) \\
LCLG.cmp.X    & 1465 & 4.5 (5.5)& 15.5 (25.5)& 0.46 (0.77)& 180 (265)& 13.25 (13.75)&11.15 (11.45) \\
&&&&&&& \\
LCLG.APM      &   85 & 4.5 (7.5)& 14.5 (26.0)& 0.55 (0.81)& 210 (295)& 13.55 (13.90)&11.15 (11.55) \\
LCLG.cmp.APM  & 1377 & 4.5 (5.5)& 16.5 (24.5)& 0.46 (0.77)& 180 (265)& 13.25 (13.75)&11.15 (11.45) \\

\hline
\end{tabular}
\label{tab:med}

{\it Endnotes:}
The columns are as follows:

\noindent Column (1):
Sample identification, given in Sect. 3.

\noindent Column (2): 
$N_{\rm group}$, the number of LCLGs in the sample.

\noindent Column (3): $N_{\rm obs}$, median value of 
the observed number of LCRS galaxies in groups.

\noindent Column (4): $N_{\rm ACO}$, median value of 
the group's Abell counts. 

\noindent Column (5): $R_{\rm h}$, median value of the harmonic radius
of the groups, in units of \Mpc.

\noindent Column (6): $\sigma_{\rm los}$, median value of the group 
line-of-sight velocity dispersions, in units of km~s$^{-1}$. 

\noindent Column (7): $ M_{\rm vir}$, median value of 
the  group's virial mass, in units of $h^{-1}~M_{\sun}$.

\noindent Column (8): $L_{\rm tot}$, the median value of 
the total group luminosity in the LCRS $R$-band, in
units of solar luminosity ($h^{-2}~L_{\sun}$).

\end{table*}

\section{Properties of LCLGs in the vicinity of rich clusters}

Next, we compare the properties of LCLGs in the cluster environment
with the average properties calculated for the comparison samples for
each sample under study.

In the LCLG catalogue several physical properties have been calculated
for each group (TUC, Sect. 4).  These include the observed number of
group member galaxies $N_{\rm obs}$, the harmonic radius $R_{\rm h}$,
the line-of-sight velocity dispersion $\sigma_{\rm los}$, the virial mass
$M_{\rm vir}$, the total luminosity $L_{\rm tot}$, and the Abell
counts $N_{\rm ACO}$.

\begin{figure*}
\vspace{5cm}
\caption{The distribution of the observed group membership 
(observed numbers of galaxies) for LCLGs near a  rich cluster.  In all
three panels, the {\em dashed line\/} represents the 
distribution for groups from comparison sample (see text;
in the middle panel the comparison sample for loose groups 
around X-ray clusters is not plotted since it almost coincides with 
the comparison sample for Abell clusters).  
{\bf Left panel:} Solid line: LCLGs
around  Abell-class LCLGs ($N_{ACO} \geq 30$). 
{\bf Middle panel:} LCLGs around Abell clusters 
({\em thin solid line\/}) and X-ray
clusters ({\em bold solid line\/}).  
{\bf Right panel:} LCLGs around APM
clusters (the {\em solid line\/}).
 }
\includegraphics{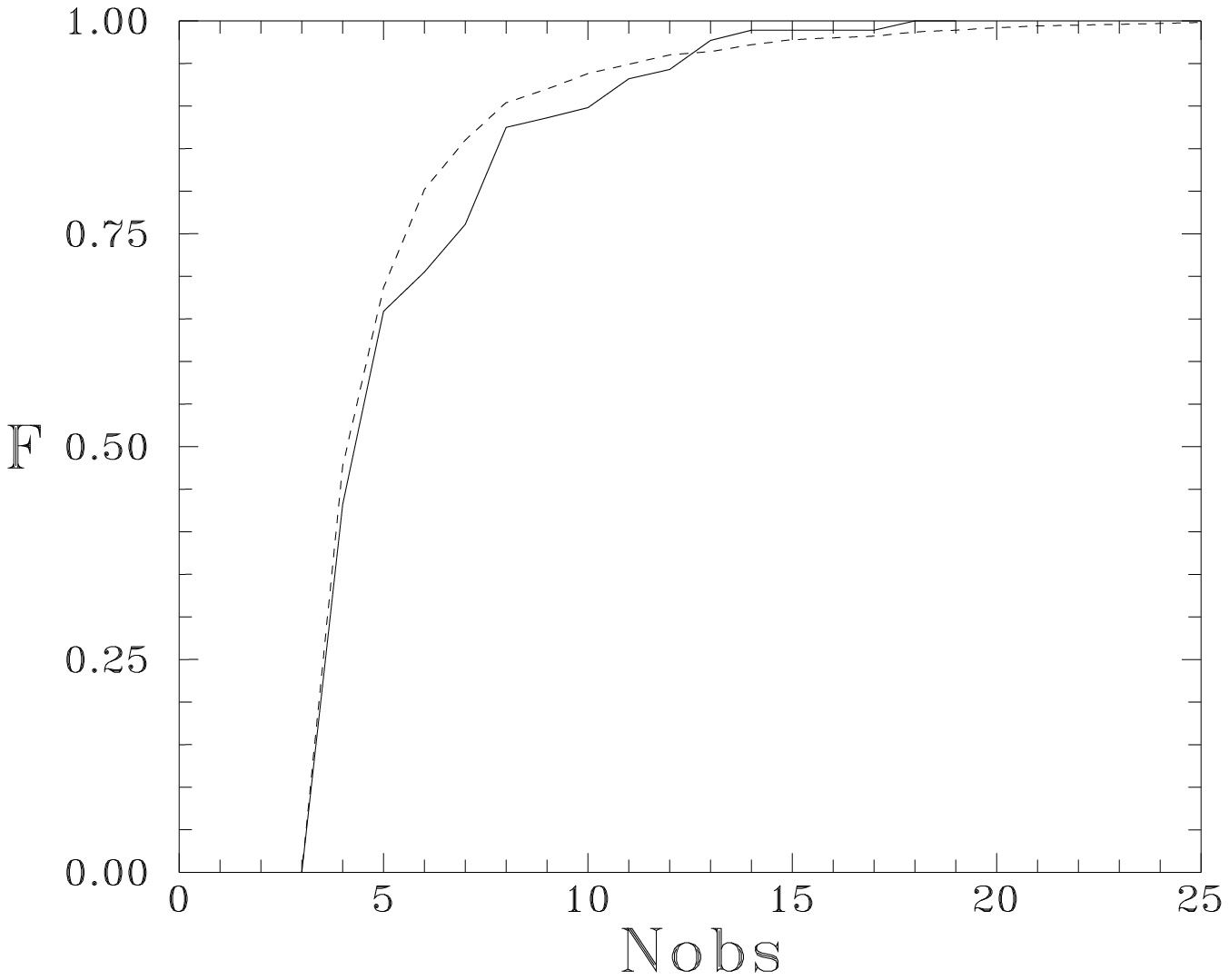}
\includegraphics{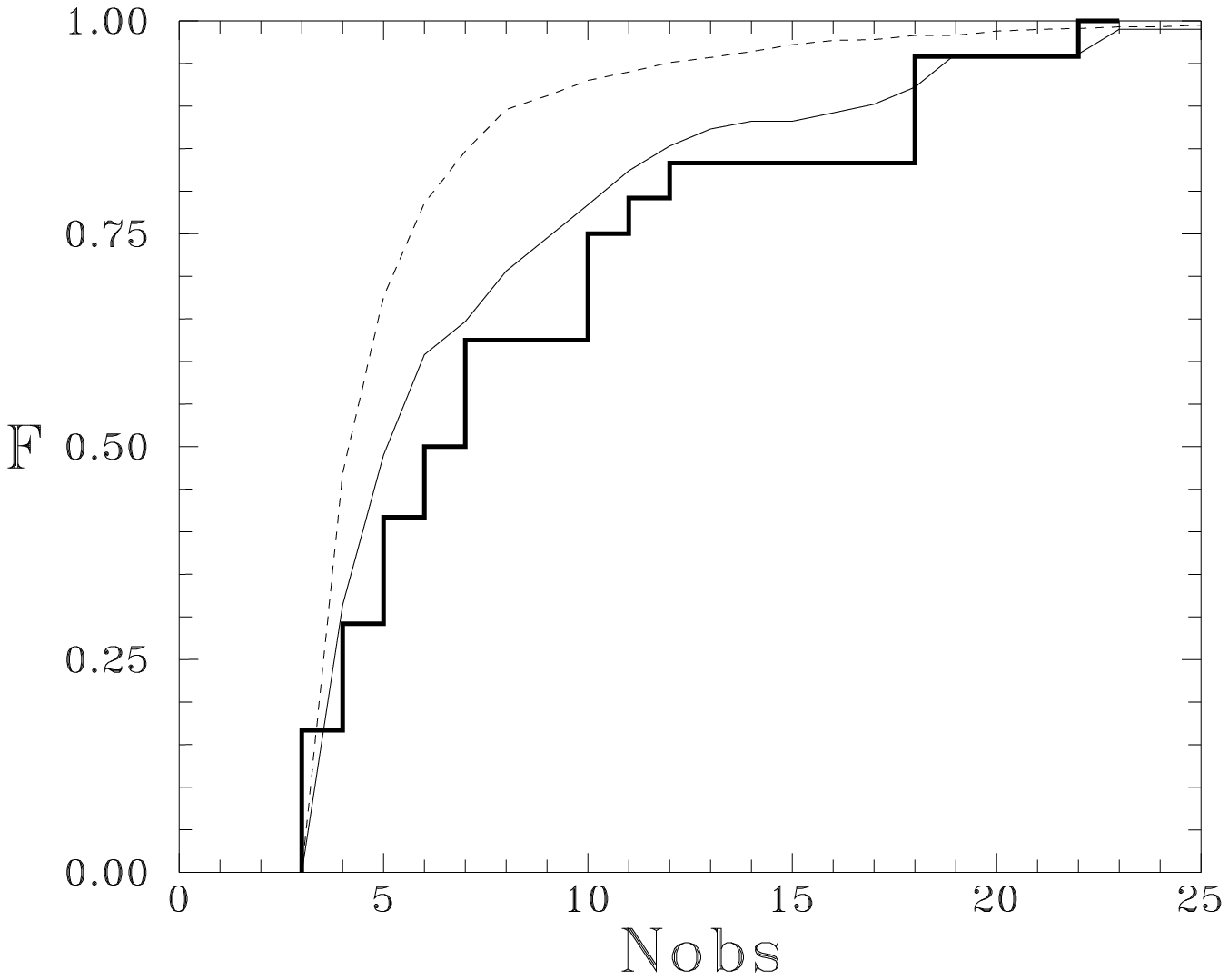}
\includegraphics{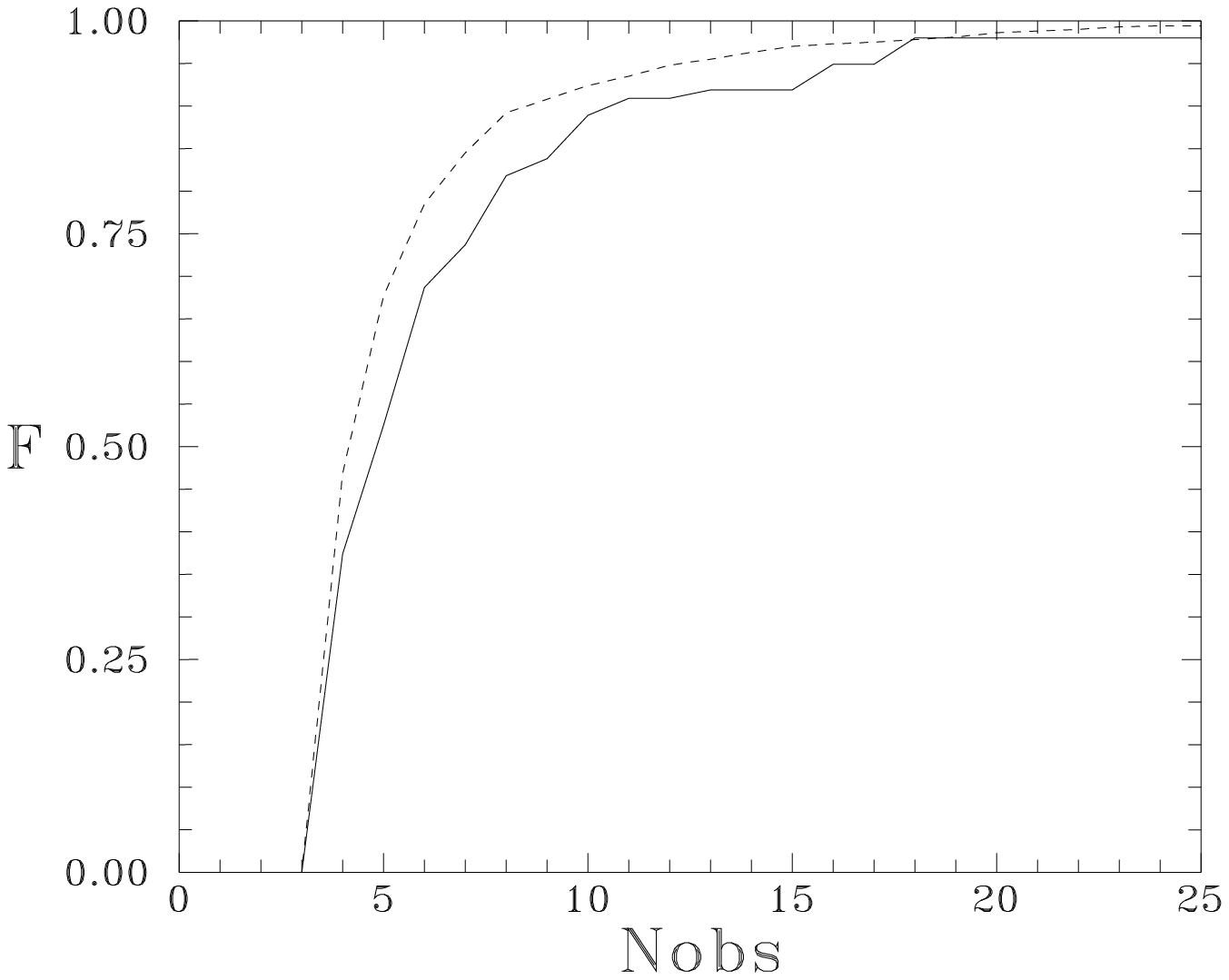}
\label{fig:lgno}
\end{figure*}

\begin{figure*}
\vspace{5cm}
\caption{Distribution of Abell counts for LCLGs around rich clusters.
Panels and lines are as in 
Fig.~\ref{fig:lgno} }
\includegraphics{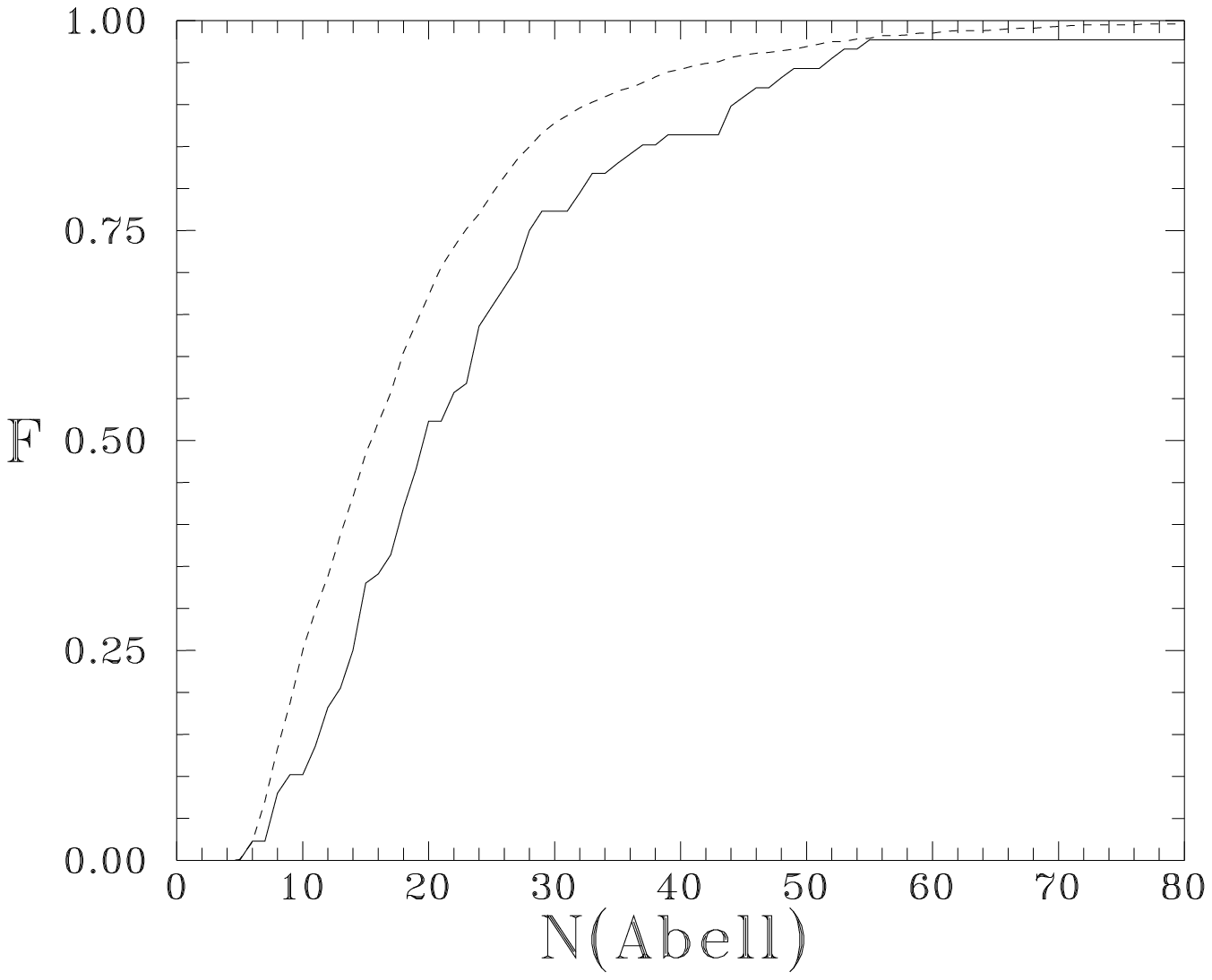}
\includegraphics{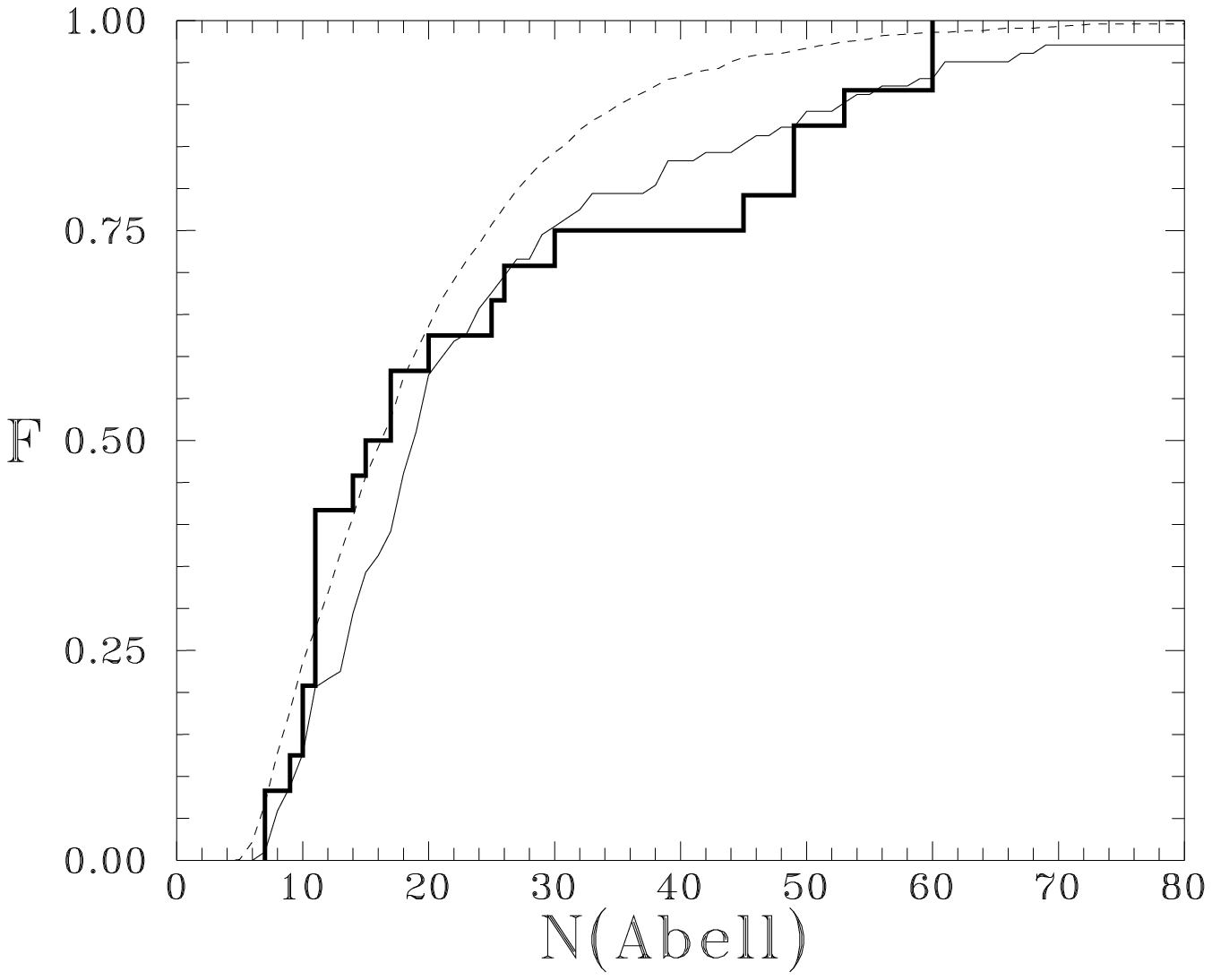}
\includegraphics{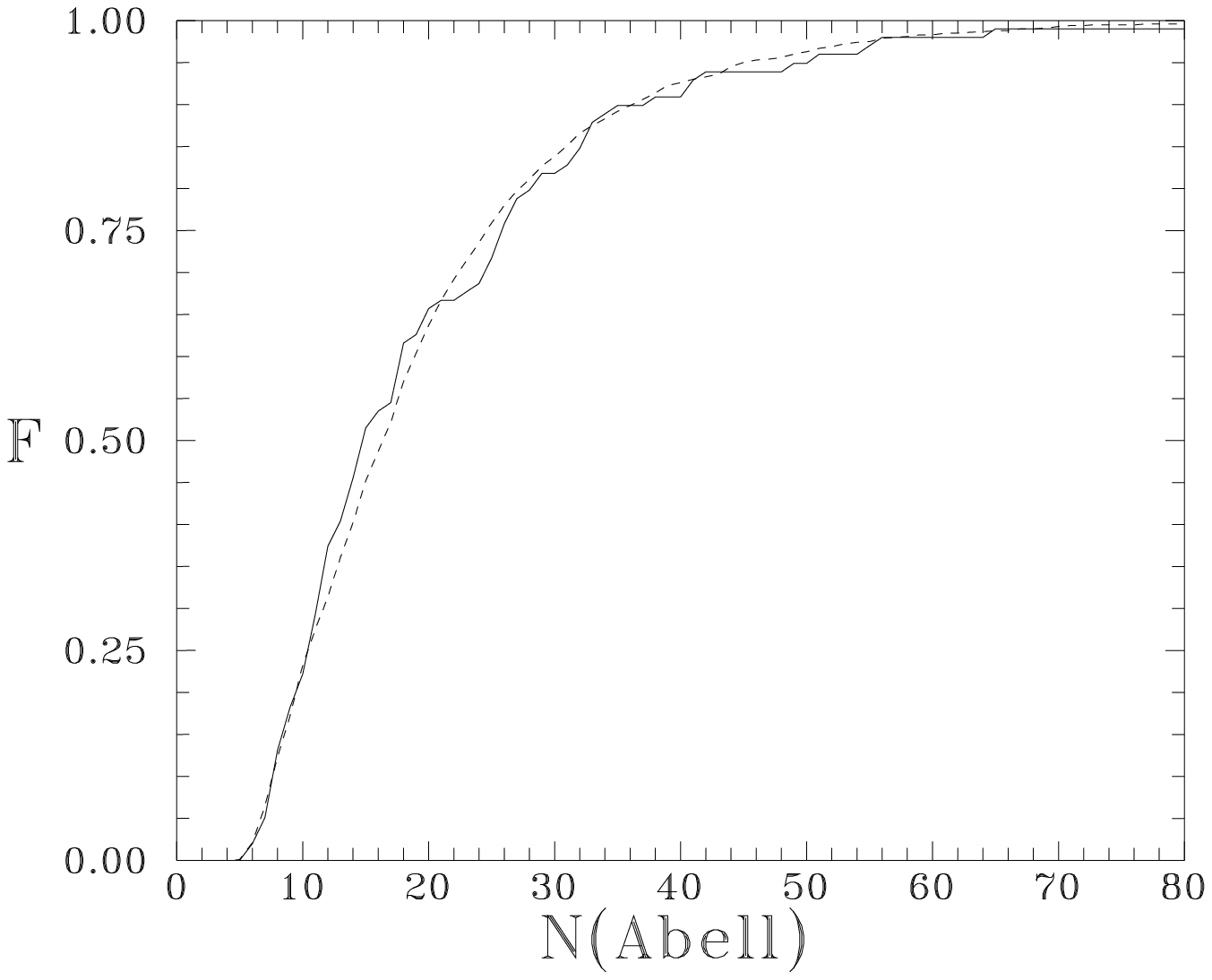}
\label{fig:lgna}
\end{figure*}

\begin{figure*}
\vspace{5cm}
\caption{Distribution of harmonic radii (in \Mpc) for LCLGs around rich
clusters.  Panels and 
lines are as in Fig.~\ref{fig:lgno} }
\includegraphics{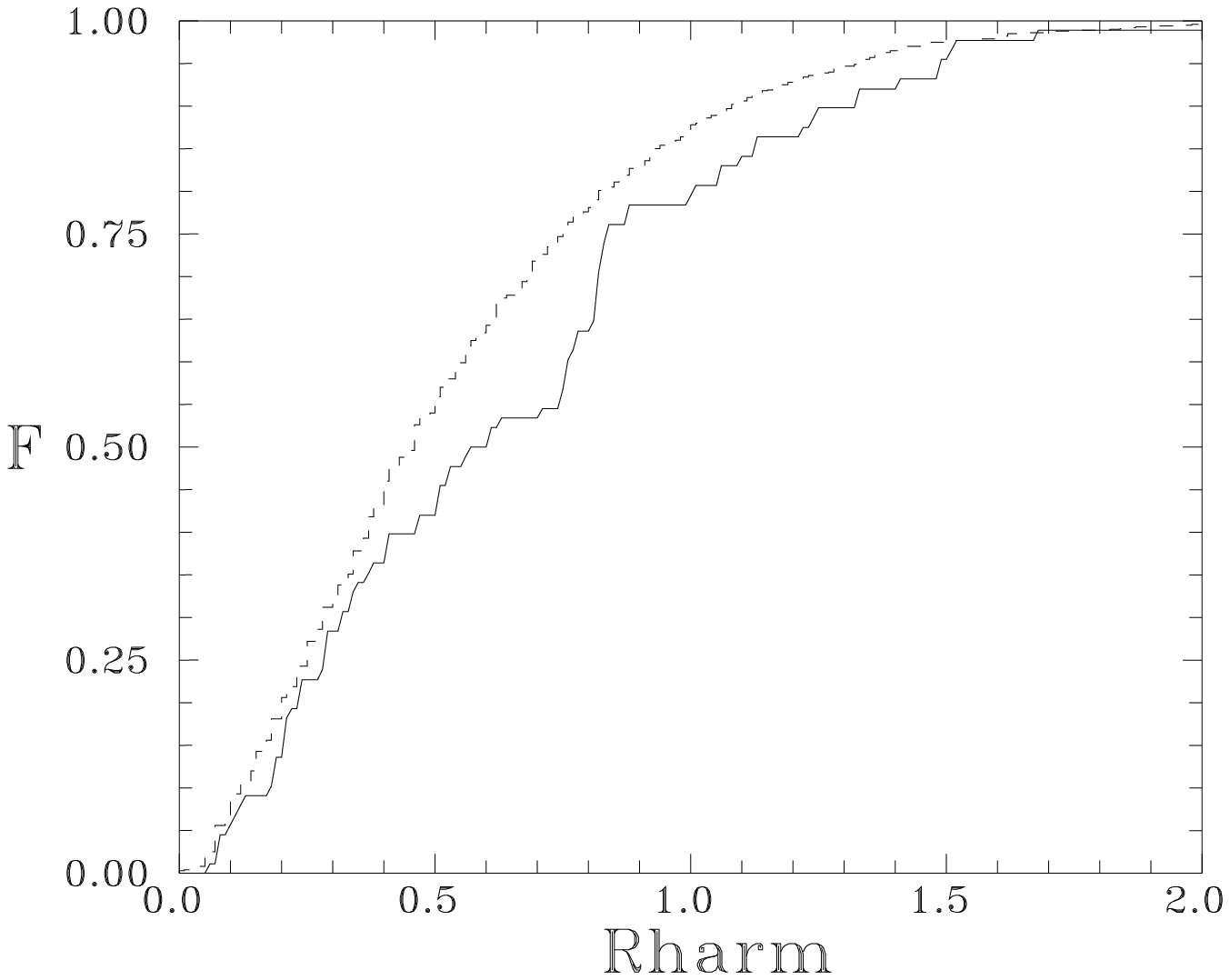}
\includegraphics{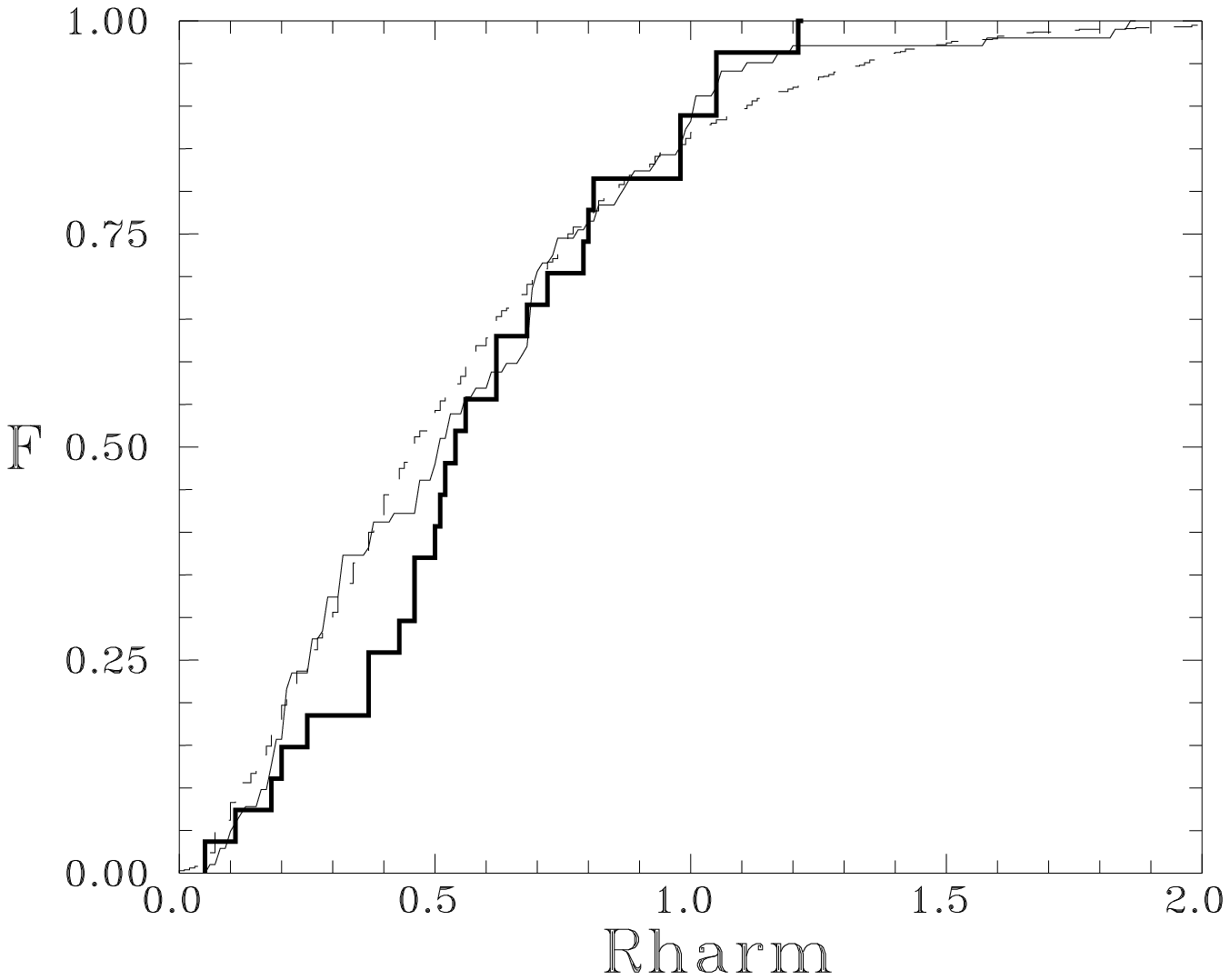}
\includegraphics{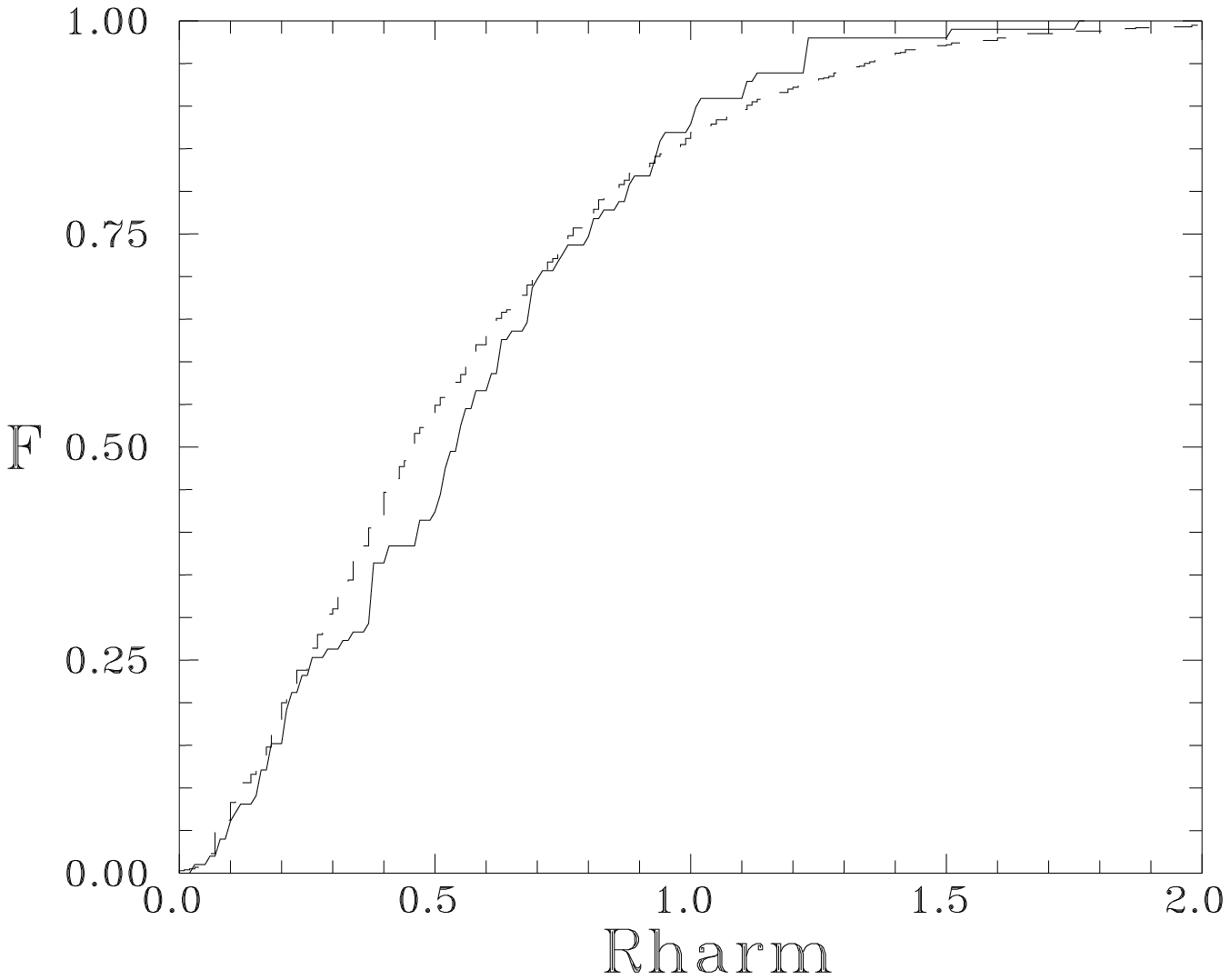}
\label{fig:lgr}
\end{figure*}

\begin{figure*}
\vspace{5cm}
\caption{Distribution of velocity dispersions, $\sigma_{\rm los}$ (km/s) 
for LCLGs around rich
clusters.  Panels and lines are as in Fig.~\ref{fig:lgno} }
\includegraphics{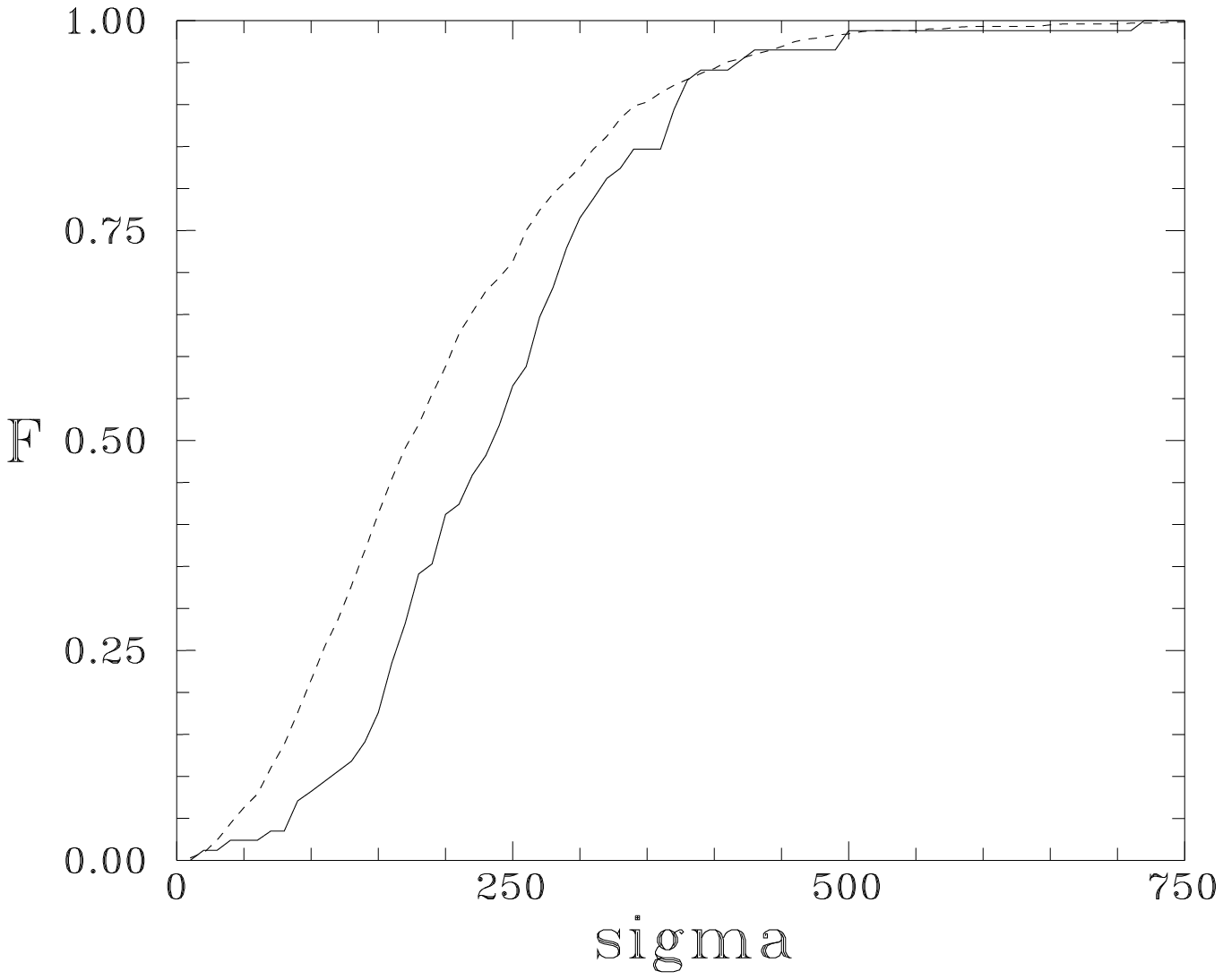}
\includegraphics{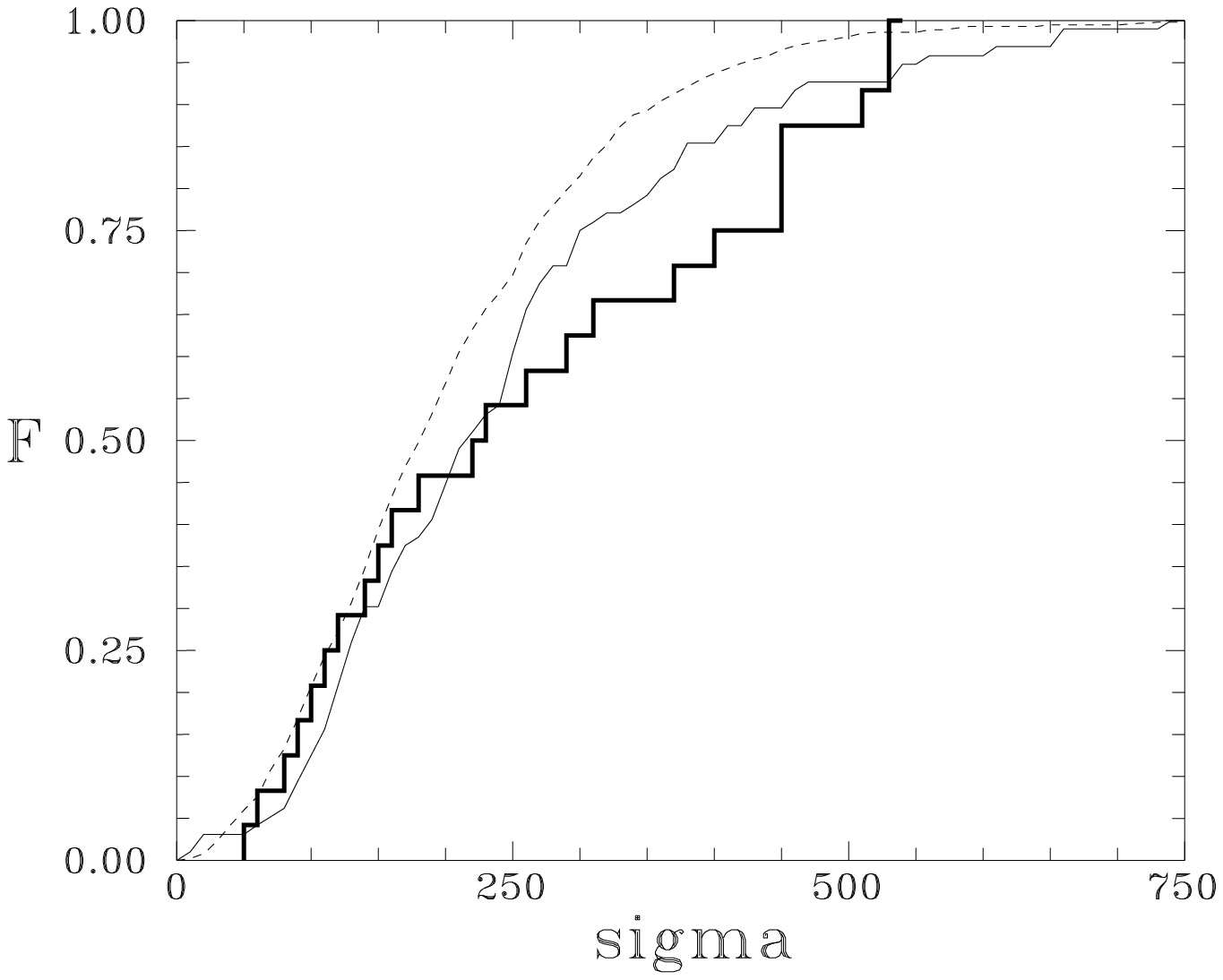}
\includegraphics{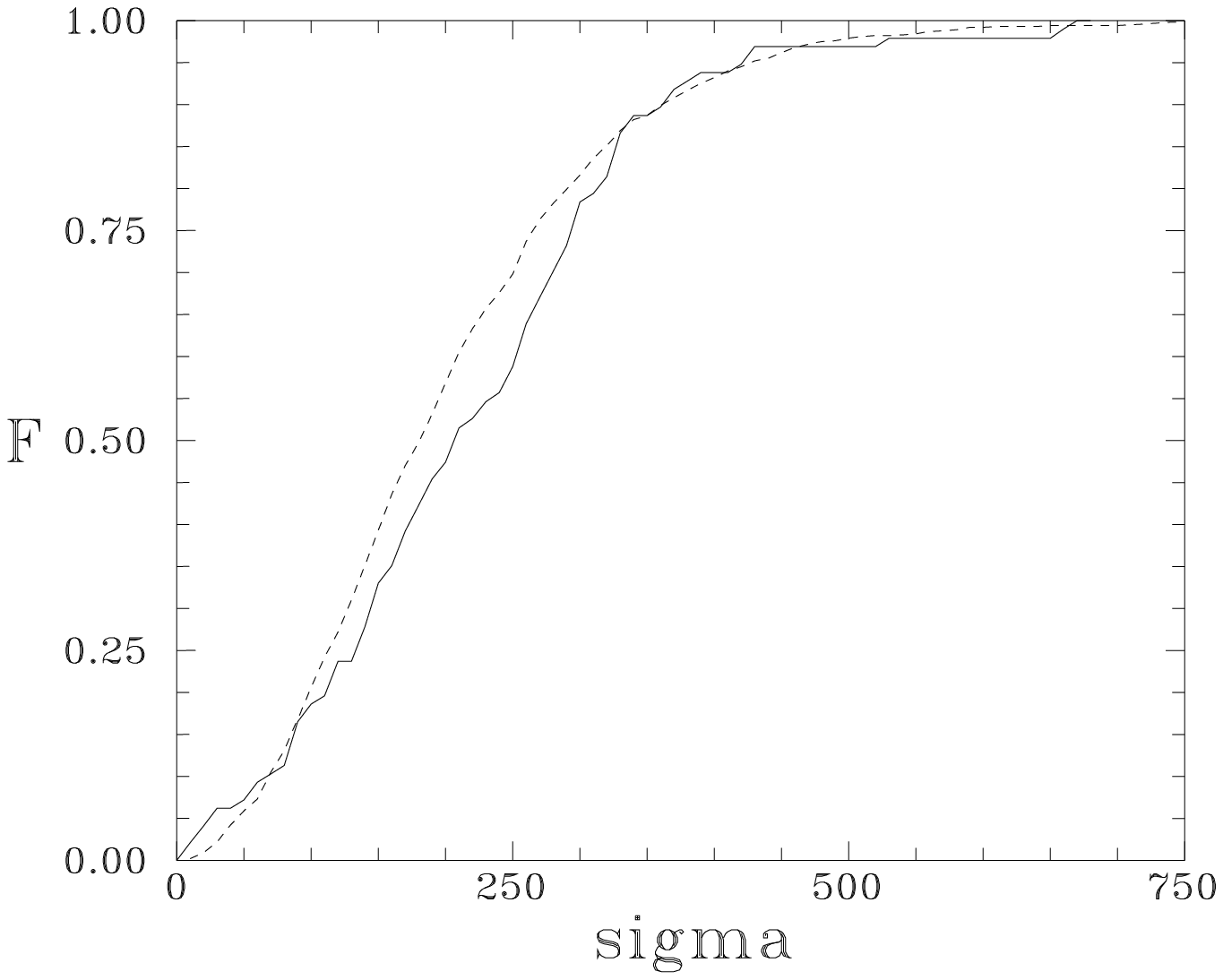}
\label{fig:lgv}
\end{figure*}

\begin{figure*}
\vspace{5cm}
\caption{Distribution of masses ($h^{-1}~M_{\sun}$) 
for LCLGs around rich clusters.  
Panels and lines are as in Fig.~\ref{fig:lgno} }
\includegraphics{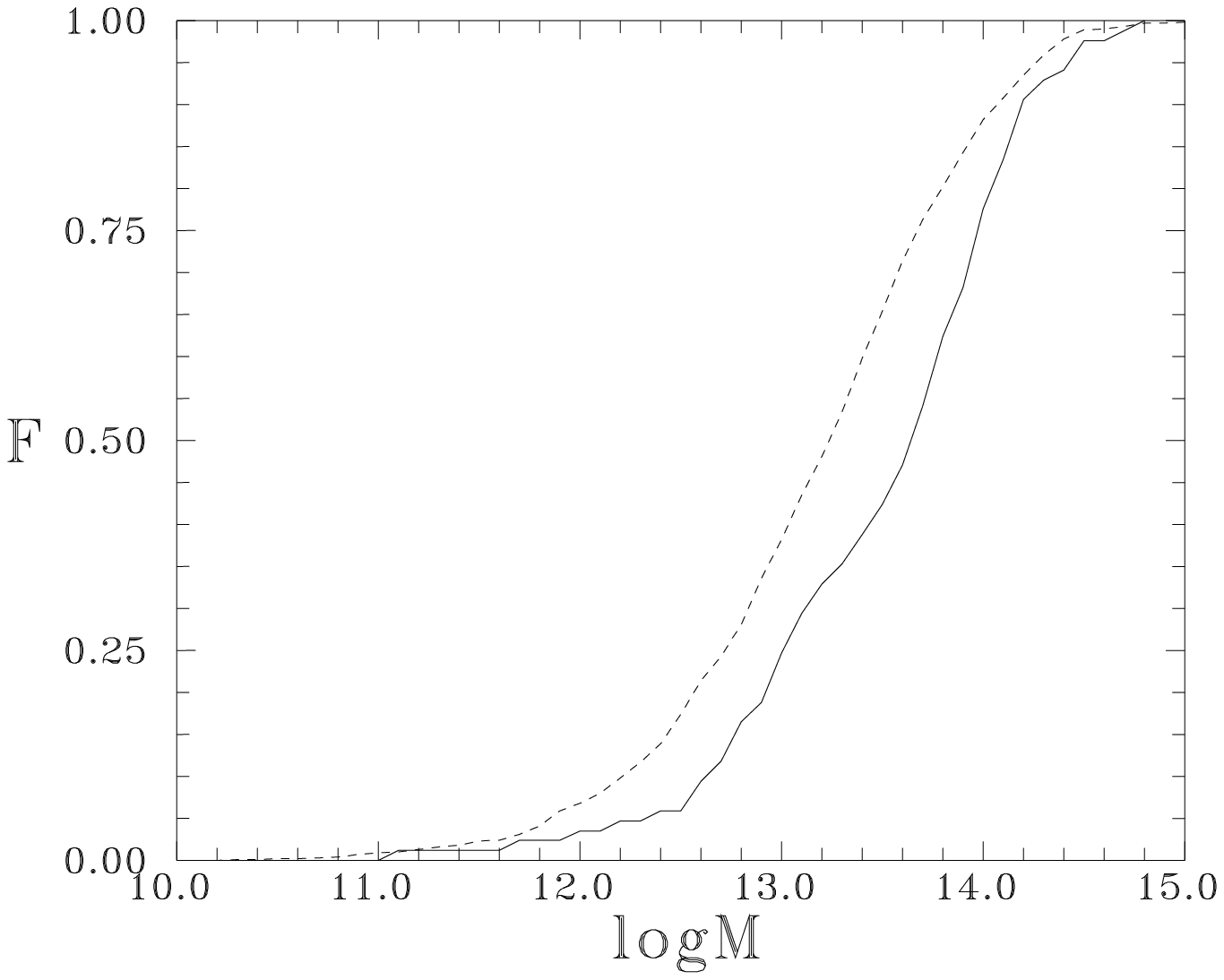}
\includegraphics{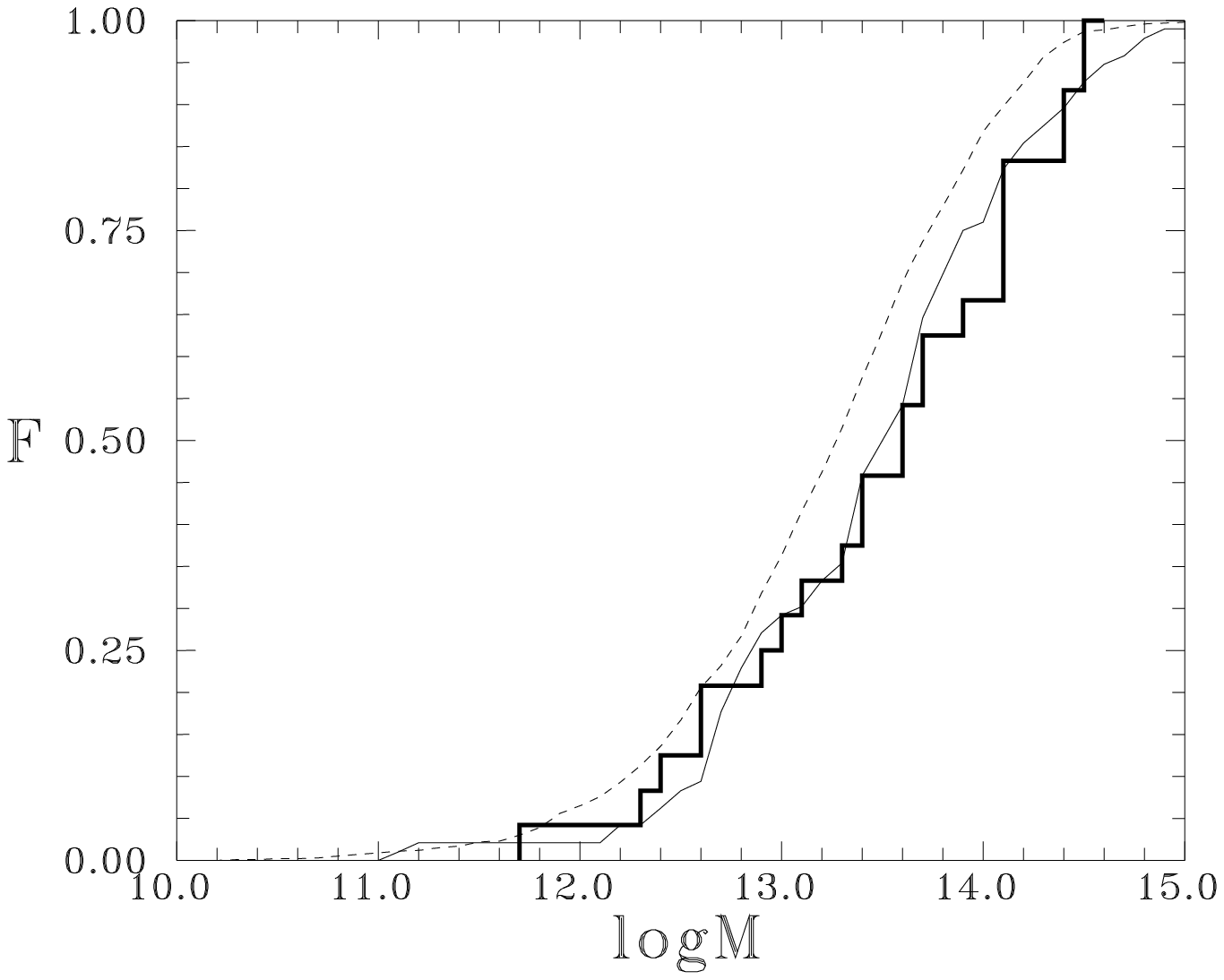}
\includegraphics{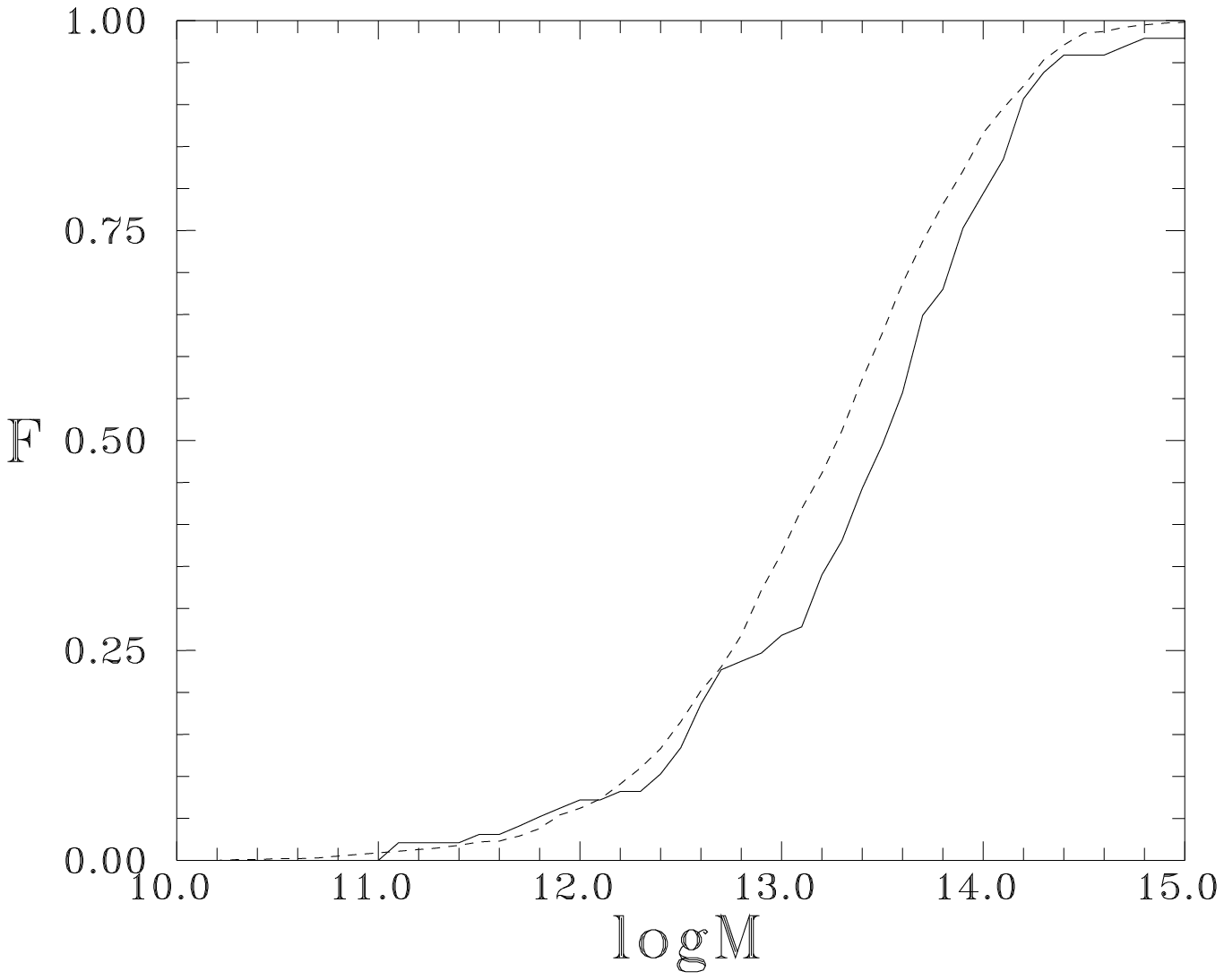}
\label{fig:lgm}
\end{figure*}

\begin{figure*}
\vspace{6cm}
\caption{Distribution of luminosities 
(in units of solar luminosity $h^{-2}~L_{\sun}$)
for LCLGs around rich clusters.
Panels and lines are as in Fig.~\ref{fig:lgno} }
\includegraphics{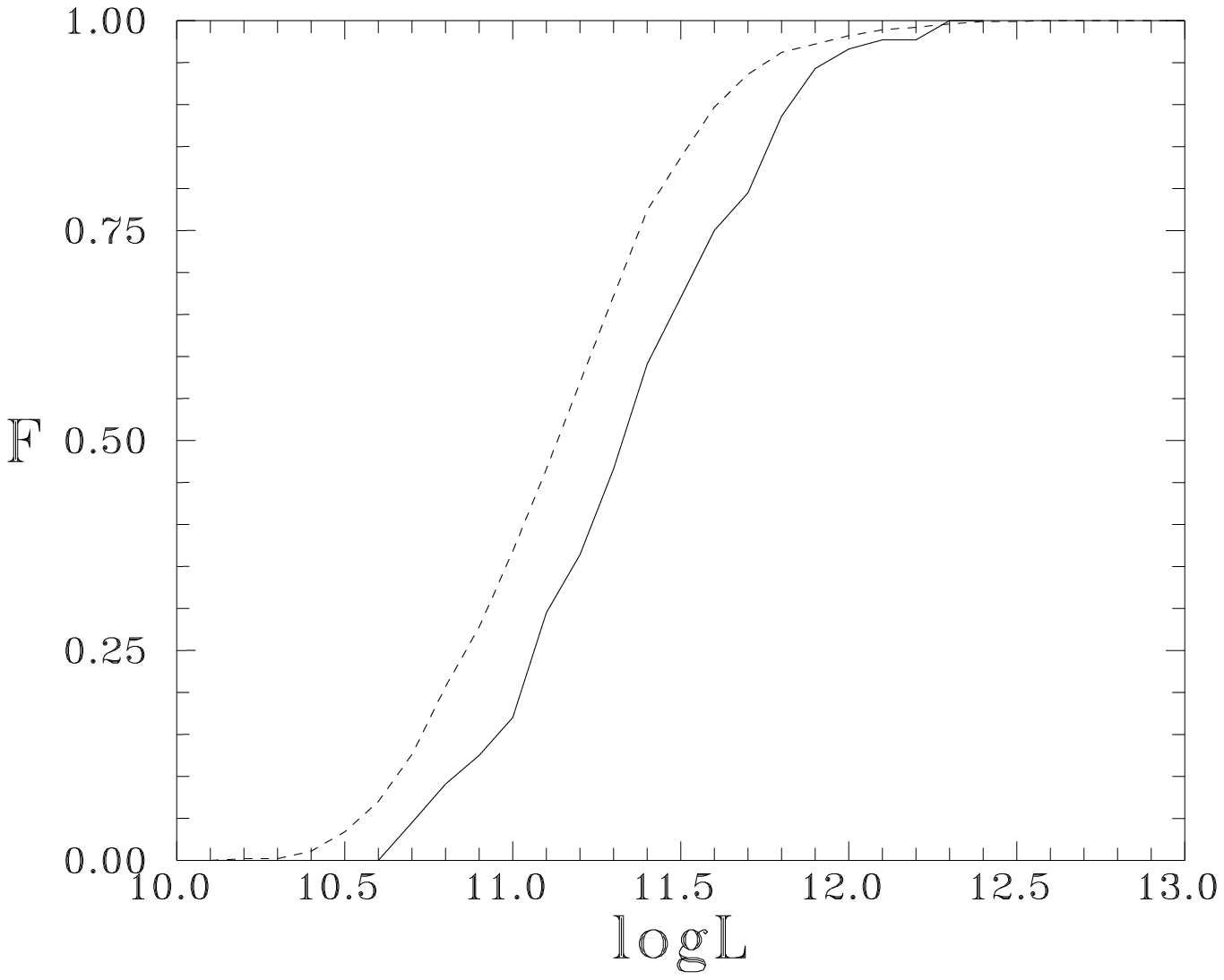}
\includegraphics{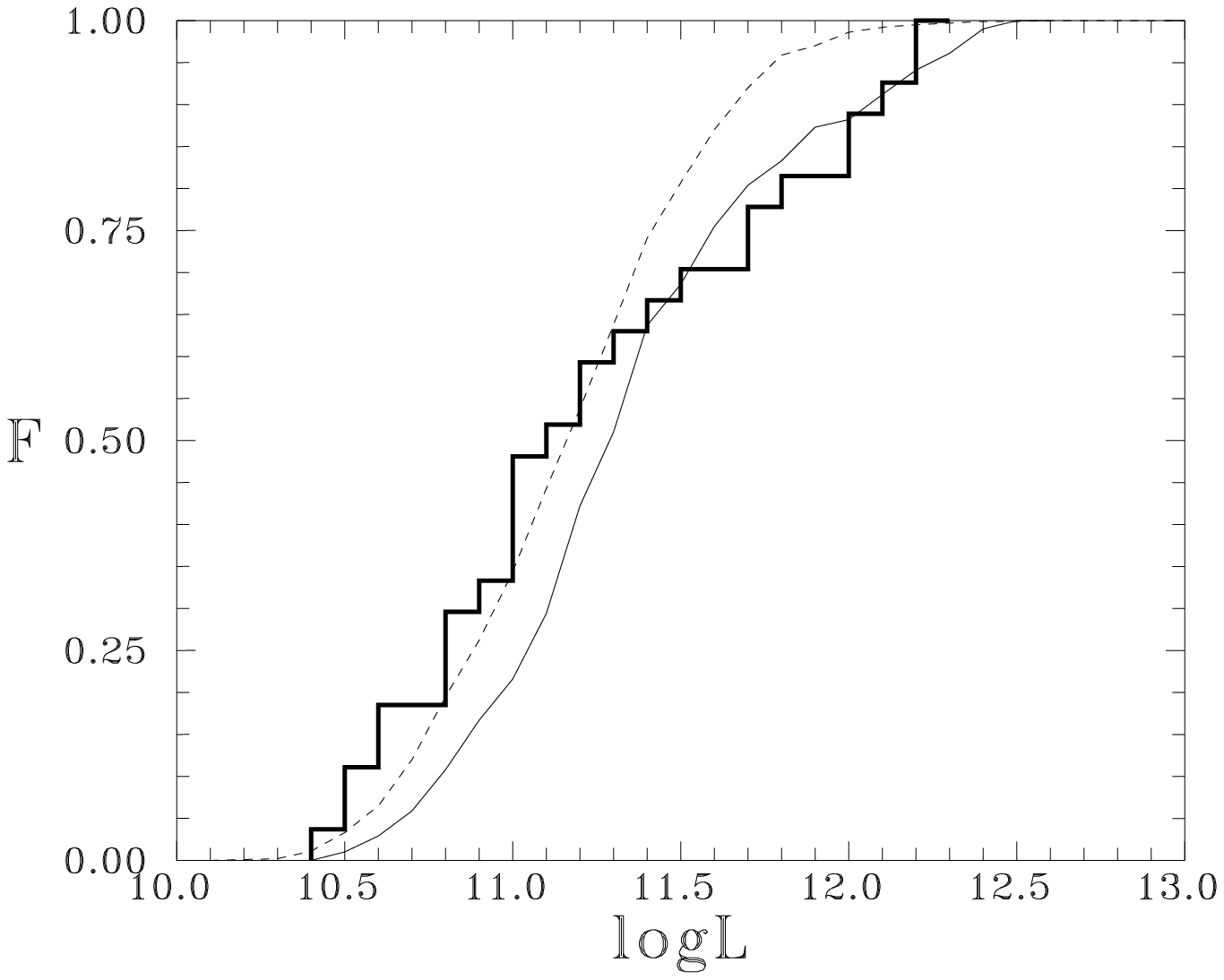}
\includegraphics{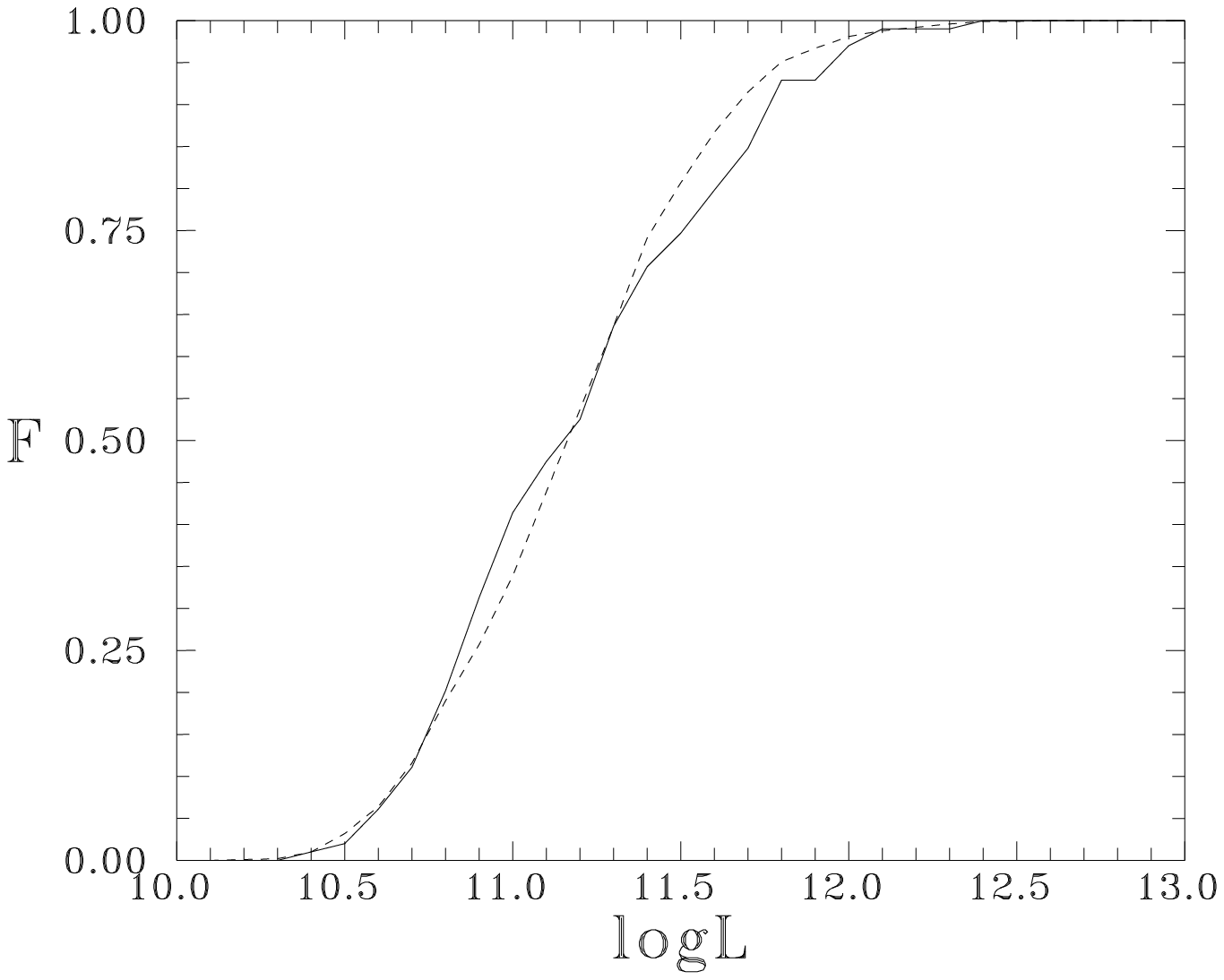}
\label{fig:lgl}
\end{figure*}

Each of these properties highlights a different physical aspect of
groups.  Thus, it is useful to analyse how several of these properties
differ between a sample of groups in high-density environments and for
the comparison sample. We refer to TUC for details of how these properties
were estimated for the LCLG catalogue.
Now let us analyse the properties of loose groups in more detail.

The results of our calculations are shown in Table~\ref{tab:med},
where we give the median and upper quartile values of the properties
of groups from various samples. The cumulative distributions of the
properties of loose groups are plotted in Figs.~\ref{fig:lgno} --
\ref{fig:lgl}.  We have also estimated the statistical significance of
our results with the Kolmogorov-Smirnov test (Table~\ref{tab:ks}).

First, consider the distribution of observed group richness values (number
of galaxies) shown in Fig.~\ref{fig:lgno}.  In most cases, an LCLG
near a rich cluster typically has a larger observed group membership
than loose groups from the comparison catalogue. Note that the
differences between the test and the comparison samples are larger for
the upper quartiles (Table~\ref{tab:med}).  

The observed membership of loose groups, however, is fraught with
selection effects.  A better measure of a group's richness is its
Abell count $N_{\rm ACO}$.  To first order, the measured Abell counts
for LCLGs should be independent of selection effects (see TUC).  When
using the Abell counts as the measure of group richness, a LCLG near
a rich cluster tends to be richer than a typical loose group drawn
from the comparison sample (Fig.~\ref{fig:lgna}).

Figure~\ref{fig:lgr} shows that the harmonic radii of loose groups in
the neighbourhood of rich clusters are somewhat larger than those of
loose groups on average in the comparison sample. The largest
differences occur between loose groups around  Abell-class LCLGs and the
smallest differences 
in the case of loose groups around Abell and APM clusters.

Figure~\ref{fig:lgv} shows that the velocity dispersion of loose groups
in the neighbourhood of rich clusters is about $1.3$ times larger than
that in the comparison samples (these groups are ``hotter'').  The
upper quartile values for velocity dispersions of groups in the
neighbourhood of X-ray clusters are especially large.

Figure~\ref{fig:lgm} shows the distribution of masses of loose groups in
the vicinity of rich clusters.  This figure shows the presence of a
strong mass segregation -- loose groups in the vicinity of rich
clusters have masses that are, on average, about $2.5$ times larger
than mean masses of loose groups.

Figure~\ref{fig:lgl} shows that loose groups in the neighbourhood of
 Abell-class loose groups and in the neighbourhood of
Abell clusters are $1.6$ times more
luminous than loose groups from comparison samples. In the case of
loose groups around X-ray clusters the upper quartile values show a
large difference between dense-environment groups and  the comparison
sample. The smallest differences are found for loose groups around APM
clusters.

Next let us check whether the properties of loose groups around rich
clusters depend on the properties of the rich clusters themselves.

\begin{figure}
\vspace{6cm}
\caption{Masses of LCLGs around Abell clusters (sample LCLG.Abell, 
masses are in units of $h^{-1}~M_{\sun}$), 
against the richnesses of these Abell clusters.
}
\includegraphics{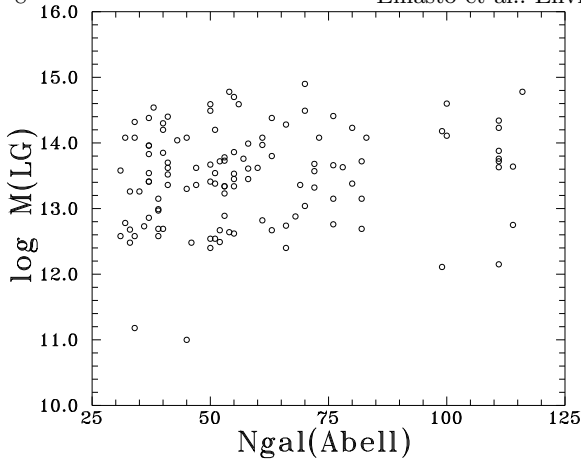}
\label{fig:ngala6m}
\end{figure}

First, we consider the richness of Abell and APM clusters.  In
Fig.~\ref{fig:ngala6m} we plot the masses of loose groups around
Abell clusters against the Abell cluster richnesses (the results for
APM clusters are similar, but to avoid overcrowding they are
not shown).  Figure~\ref{fig:ngala6m} shows no clear dependence on the
cluster's richness. We obtain a similar result if we plot group
properties other than mass.

\begin{figure}
\vspace{6cm}
\caption{Masses of LCLGs around  Abell-class loose groups
(sample LCLG.LG), against the masses of these  Abell-class groups
(in units of $h^{-1}~M_{\sun}$).
}
\includegraphics{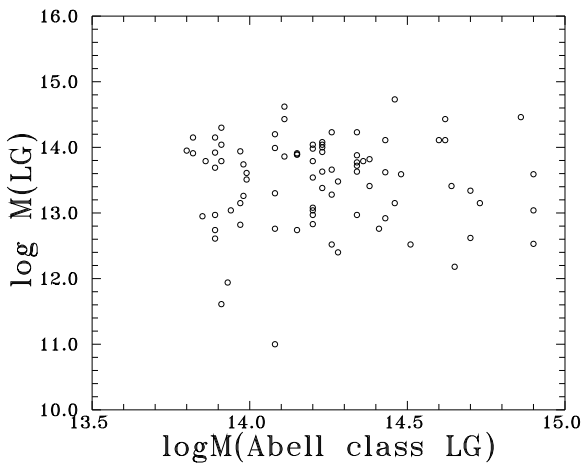}
\label{fig:lgcm}
\end{figure}

In Fig.~\ref{fig:lgcm} we show the masses of loose groups around
Abell-class loose groups against the masses of these Abell-class loose
groups themselves.  Again, this figure shows no clear dependence of a
loose group's mass on the mass of the nearby  Abell-class group.

Summarising, these figures demonstrate a strong environmental
enhancement of the mass and the richness of loose groups in the
vicinity of rich clusters of galaxies.  The differences between
properties of loose groups in the neighbourhood of rich clusters and
those of loose groups away from rich clusters are minimal in the
case of loose groups around APM clusters.  This may be due to the fact
that APM clusters are themselves poorer than rich clusters from the
other samples.  
Somewhat lower confidence levels in the statistical
significance of the differences between samples in the case of loose
groups around X-ray clusters (Table~\ref{tab:ks}) are likely due to
the small number of loose groups in this sample.  Finally, we note that, 
although loose group properties do appear to depend on proximity
to rich clusters, they do not appear to depend on the richness of
rich clusters themselves.

\begin{table*}
\caption{Confidence levels of the statistical significance of 
the Kolmogorov-Smirnov tests}

\begin{tabular}{lrrrrrrr}
Sample &   $N_{\rm group}$   &   $N_{\rm obs}$ & $N_{\rm ACO}$ & $R_{\rm h}$ &  $\sigma_{\rm los}$ & $\log M_{\rm vir}$ & $\log L_{\rm tot}$  \\
 (1)   &      (2)          &        (3)      &     (4)       &     (5)     &          (6)        &         (7)        &         (8)         \\

\hline 
LCLG.LG       &  95    & 75\% & 99\% & 99\% & 98\% & 99\% & 99\% \\
LCLG.Abell    &  96    & 99\% & 90\% & 75\% & 75\% & 97\% & 95\% \\
LCLG.X        &  24    & 99\% & 75\% & 70\% & 85\% & 80\% & 70\% \\
LCLG.APM      &  85    & 90\% & 65\% & 75\% & 90\% & 90\% & 65\% \\

\hline
\end{tabular}
\label{tab:ks}

{\it Endnotes:}
The columns are as follows:

\noindent Column (1): Sample identification, given in Sect. 3.

\noindent Column (2): $N_{\rm group}$, the number of LCLGs in the sample.

\noindent Column (3): KS statistical significance for $N_{\rm obs}$.

\noindent Column (4): KS statistical significance for $N_{\rm ACO}$.

\noindent Column (5): KS statistical significance for $R_{\rm h}$.

\noindent Column (6): KS statistical significance for $\sigma_{\rm los}$.

\noindent Column (7): KS statistical significance for $M_{\rm vir}$.

\noindent Column (8): KS statistical significance for $L_{\rm tot}$.

\end{table*}

\section{Discussion and conclusions}

We have analysed the properties of loose groups from the LCRS in high
density regions as determined by the location of rich optical and
X-ray clusters.  Our study demonstrates an environmental enhancement
of some characteristic properties of these groups: they are richer,
more massive, and more luminous than typical loose groups on average.
Now we shall discuss the confidence of our results. First we
consider selection effects.

\subsection{The properties of loose groups  and 
selection effects}

There are two main selection effects in the LCRS: in apparent
magnitude, and an almost random selection due to the limited number of
fibres used in the survey [see also TUC, Doroshkevich et al. (\cite{dor01}), 
and Hein\"am\"aki et al. (\cite{hei:hei})].
 
\begin{figure}
\vspace{7cm}
\caption{Luminosities of LCRS loose groups from total
sample of the LCLGs (dots) and from the sample 
of groups near  Abell-class loose groups (open circles)
(in units of solar luminosity $h^{-2}~L_{\sun}$)
against group's distances (in \Mpc\ ).
}
\includegraphics{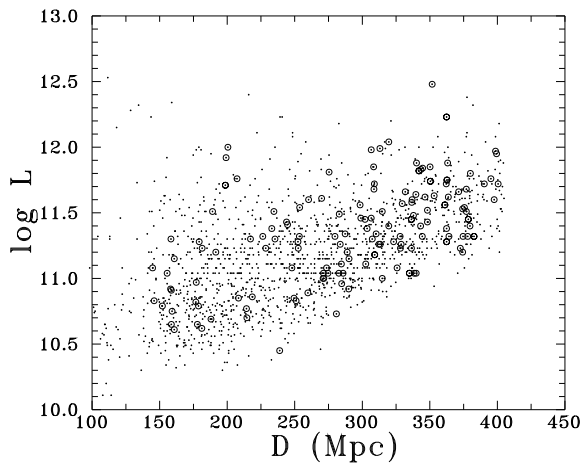}
\label{fig:lgldis}
\end{figure}

LCRS observations were performed in a fixed apparent magnitude
interval and therefore the absolute magnitude interval of observed
galaxies depends on the distance of the galaxy -- galaxies fainter or
brighter than this interval are not included in the survey. Thus
groups consisting of faint galaxies occur only in the nearest regions
of the survey. With increasing distance, groups containing fainter
galaxies gradually disappear from the sample.  This effect is seen in
Figure~\ref{fig:lgldis} where we plot the total luminosity of loose
groups against distances of these groups for two samples: the total
sample of LCLGs and the sample containing only those loose groups
around  Abell-class groups (for other samples of loose groups in the
environment of rich clusters the distribution is similar).

This figure shows that this effect is similar both for loose
groups around rich clusters and for the total sample of loose
groups. However, galaxies in low density environments tend to be fainter
(Hamilton \cite{ham88}, Einasto \cite{ein91}, Lindner et
al. \cite{lind95}, Norberg et al.  \cite{nor01}, Zehavi et
al. \cite{zeh02}), and thus at a given distance the probability that a
galaxy may not be observed in the LCRS window of apparent magnitudes is
higher in low density regions. Thus loose groups in the high density
environments studied in the present paper are probably less affected by
the luminosity selection effect than loose groups on average.

However, Figure~\ref{fig:lgldis} shows that there are no loose 
groups in the neighbourhood of rich clusters among the nearest 
groups included in the comparison sample.
Could it be possible that the environmental
effect we study in the present paper is simply a selection 
effect due to the luminosity-distance relation? 

\begin{figure}
\vspace{7cm}
\caption{Distribution of masses ($h^{-1}~M_{\sun}$) 
for LCLGs around  Abell-class loose groups 
(bold lines), and for loose groups from comparison
samples (thin lines).  
Solid lines: samples from distance interval 150 - 400 \Mpc, 
short-dashed lines: samples from distance interval 200 - 400 \Mpc,
and long-dashed lines: samples from distance interval 250 - 400 \Mpc.
}
\includegraphics{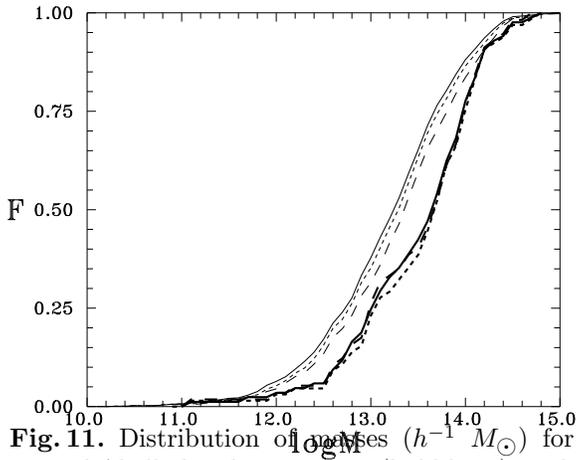}
\label{fig:lg1525m}
\end{figure}

To clarify this problem we choose both samples of 
loose groups around rich clusters and comparison samples from
three distance intervals: $150 - 400$~\Mpc, $200 - 400$~\Mpc,
and $250 - 400$\Mpc.
Then we derived the distribution 
of the properties of these distance-limited samples. In
Table~\ref{tab:dist} we give the median and upper quartile
values of the properties of groups from these distance intervals
in the case of loose groups around  Abell-class groups (sample LCLG.LG). 
Figure~\ref{fig:lg1525m} shows the cumulative distribution of masses 
of groups for samples of loose groups around 
 Abell-class loose groups, and for corresponding comparison samples
from different distance intervals. In the case of loose groups
around other rich clusters the results are similar.

Table~\ref{tab:dist} and Figure~\ref{fig:lg1525m}
show clearly that the selection effect due to the 
luminosity-distance relation does 
not change our main conclusion:
properties of groups in the neighbourhood of 
rich clusters are enhanced in comparison with the properties of
loose groups on average.

The  Kolmogorov-Smirnov test shows that the differences
between the samples are statistically significant. 
For example, the distributions of
masses for samples from distance interval 
$250 - 400$\Mpc\ are different at the 95\% confidence level.

\begin{table*}
\caption{Median and upper quartile (in parentheses) values of LCLG properties.}

\begin{tabular}{lrrrrrrr}
Sample & $N_{\rm group}$ & $N_{\rm obs}$ & $N_{\rm ACO}$ & $R_{\rm h}$
&  $\sigma_{\rm los}$ 
& $\log M_{\rm vir}$ & $\log L_{\rm tot}$  \\
& & & &\Mpc\ 
&   km~s$^{-1}$ & $h^{-1}M_{\odot}$  & $h^{-2}L_{\odot}$\\
(1)& (2) & (3) & (4) & (5)& (6) &(7) & (8)  \\

\hline 
LCLG.LG.150         & 95 & 4.5 (7.0)& 19.5 (28.0)& 0.65 (0.89)& 235 (295)& 13.65 (14.00)&11.35 (11.65) \\
LCLG.LG.200         & 65 & 4.5 (7.0)& 23.5 (32.5)& 0.74 (0.99)& 235 (280)& 13.65 (14.00)&11.35 (11.65) \\
LCLG.LG.250         & 57 & 4.5 (7.0)& 23.5 (32.5)& 0.62 (1.05)& 225 (290)& 13.70 (14.05)&11.40 (11.65) \\
&&&&&&& \\
LCLG.cmp.LG.150   & 1223 & 4.0 (5.5)& 16.0 (25.0)& 0.46 (0.75)& 175 (265)& 13.25 (13.70)&11.15 (11.35) \\
LCLG.cmp.LG.200   &  930 & 4.5 (5.5)& 16.5 (25.5)& 0.47 (0.80)& 185 (265)& 13.25 (13.75)&11.25 (11.35) \\
LCLG.cmp.LG.250   &  639 & 4.5 (6.0)& 17.5 (26.0)& 0.50 (0.87)& 185 (275)& 13.35 (13.80)&11.25 (11.45) \\
\hline
\end{tabular}
\label{tab:dist}

{\it The columns are as in Table~\ref{tab:med}.
The lower distance limit is shown in the name of the sample.}

\end{table*}

Another selection effect in the LCRS is due to different number of 
fibres used during redshift measurements: depending on the field, 50 
or 112 galaxies were measured for redshifts,  but the actual number of 
galaxies in the magnitude window is larger, and the sample is diluted.  
Dilution decreases not only the number of galaxies in systems
but also the number 
of groups as the number of galaxies in the group may fall below 
the  critical value of 3 used 
in the group search (see Hein\"am\"aki et al. 2002).  In high density 
environments this effect may be stronger than in low density 
environment. When compiling the list 
of loose groups, TUC attempted to take this effect into account.  
However there is no simple and obvious way to correct for this 
selection effect for very poor groups. 

Recently, Doroshkevich et al. (\cite{dor01}) analysed selection
effects in the LCRS catalogue and concluded that selection effects are
stronger in low density environments; in high density environments these
effects may serve merely to decrease the observed richnesses of galaxy
systems.

\subsection{The local number density test for loose groups} 

So far the populations of loose groups were selected in the high
density environments determined  by the neighbourhood of rich clusters
of galaxies.  Now we check the possibility that loose groups in all
high density regions (and not just in the neighbourhood of rich
clusters) are richer, have higher velocity dispersions and so on. For
that we select loose groups from high density environments as
determined by the number density of loose groups themselves.

The  local density around loose groups can  be determined  by the number 
density of neighbours around loose groups in a spheres of a given 
radius. We searched for close neighbours for each loose group 
(independently of its richness) in  spheres of increasing 
radius. In a sphere of radius of $2$~\Mpc\ 75 loose groups have 
altogether  86 neighbours. This number is comparable with the number of 
loose groups  around each  rich cluster sample. We denote this 
sample as LG.2. In a sphere of radius of $3$~\Mpc\ already 350 groups 
have altogether 388 neighbours. We denote this sample as LG.3. 
Due to the large number of neighbours already in a $3$~\Mpc\ sphere 
we did not use still larger search radii.  This actually
means that the local number density in  the case of these test
samples is even higher than in the case of loose groups
around rich clusters. 

Moreover, in our samples of loose groups in the neighbourhood 
of rich clusters, there are no neighbours with centres 
closer than $2$~\Mpc\ to the centre of a rich cluster. Thus, 
in order to generate a test sample that most closely 
resembles the samples of loose groups around
rich clusters, we used a sample of neighbours from a shell
with inner radius $2$~\Mpc\ and outer radius $3$~\Mpc. 
This sample is denoted as LG.23, and there are 302 loose groups
in this sample. 
We note that the distance distribution
of groups from these samples is close to that of loose groups 
in the neighbourhood of rich clusters. There are only 7 loose groups
with distances less than $140$~\Mpc\ among them.

Then we derived the cumulative distributions of the properties of
loose groups from the samples LG.2, LG.3 and LG.23.  In
Table~\ref{tab:nden} we give the median values of the properties of
groups from these regions of high local number density. The corresponding
values for the total sample of the LCLGs are given for comparison
(sample LCLG.cmp).  Table~\ref{tab:nden} shows that loose groups from
the highest density regions, if the local density is determined by the
number density of loose groups themselves independently of their
richness, have smaller or comparable parameters when we compare them
with parameters of loose groups on average.  For the present study the
main conclusion from this test is: the properties of loose groups in
dense aggregates of loose groups of all richnesses, including the
poorest loose groups, are comparable to those for the LCLGs  on
average, and are not enhanced as are the properties of loose groups in the
environment of rich clusters.

\begin{table}
\begin{center}
\caption{Median  values of the properties of loose groups
from samples LG.2, LG.3 and LG.23}
\begin{tabular}{lrrrr}
\hline
Sample                                    &  LG.2&  LG.3  &LG.23 & LCLG.cmp\\   
 
\hline

$N_{\rm group}$                           &  86  &  388   & 302 & 1495\\
$N_{\rm obs}$                             &  4.0 &  4.0   & 4.0 & 4\\
$N_{\rm ACO}$                             & 11.0 & 13.0   &14.0 & 16\\
$R_{\rm h}$~(\Mpc)                         & 0.37 & 0.44   & 0.46 & 0.58 \\
$\sigma_{\rm los}$~(km~s$^{-1}$)            & 145  & 170    & 185 & 164 \\
$\log M_{\rm vir}$~($h^{-1}M_{\odot}$)      & 13.00& 13.20  &13.30 & 13.20\\
$\log L_{\rm tot}$~($h^{-2}L_{\odot}$)      & 11.05& 11.10  &11.00 & 11.10\\   
\hline
\label{tab:nden}
\end{tabular}
\end{center}
\end{table}

In addition, properties of loose groups may be affected by a selection
effect related to possible spurious members which may lead to an
artificial increase of the richnesses, the velocity dispersions and so
on. The tests above do not show the presence of this effect for loose
groups from the highest density regions. We also studied the
properties of loose groups that are located in the lowest density
regions. Loose groups from the lowest density regions can be found
using the local number density around loose groups, in analogy with
the previous test. Our calculations show that some loose groups from
the lowest density regions really have very high velocity dispersions,
masses and so on -- a hint that the properties of these groups are
affected by spurious non-member galaxies. Thus, this selection effect
may change the properties of loose groups from the lowest density
environments, but not from the high density environments.

Therefore the analysis of various selection effects do not show any
indications that selections could cause the artificial enhancement of
the properties of loose groups in high density regions in the
neighbourhood of rich clusters.

However, selection effects present in the LCRS sample are complicated.
It is possible that some of them have not been fully taken into
account.  Furthermore, this sample of loose groups is probably not
complete.  Therefore, as a next step, we plan to study the properties
of galaxy groups in high density regions using the Sloan Digital Sky
Survey (York et al. \cite{yo0}).

\subsection{Comparison with other studies}

Several recent studies of the correlation function of 
nearby groups of galaxies  indicate that the properties of 
groups of galaxies in high density regions are different from the
properties of groups on average. For example, Giuricin et al. 
(\cite{giu01}), Girardi et al. (\cite{gir00}), and Merchan et al. 
(\cite{mer00}) all concluded that groups with larger velocity dispersions
and masses are more strongly clustered than groups with lower velocity 
dispersions and masses. Stronger clustering is an 
indication that these groups could be located in the high density 
regions of superclusters (Einasto et al. \cite{e97}, Tago et al. 
\cite{tag02}). 

Other studies have also shown that the properties of groups of
galaxies in superclusters may differ from the properties of groups in
low density environments. For example, groups in superclusters
may be X-ray sources (Rines et al. \cite{rin:rin}, Bardelli et
al. \cite{bar00}, Kull \& B\"ohringer \cite{kull99}).

Using an earlier version of the LCRS loose group catalogue, Hashimoto 
et al. (\cite{hash98}) have  shown that galaxies in massive 
loose groups show a smaller star formation rate than do galaxies in 
less massive loose groups. Recently Martinez et al. (\cite{mar02}) and 
Dominguez et al. (\cite{dom02}) have also shown this effect using a 
catalogue of loose groups extracted from the 2dF galaxy redshift 
survey.    They suggest that galaxy populations in more massive groups 
are more evolved than average populations due to their evolutional 
history: these galaxies are merger remnants as expected in 
hierarchical models for galaxy formation. In the future it would be 
interesting to check the possible connection between their findings 
and our results.

The studies of the properties of rich clusters of galaxies have also
shown evidence that the properties of clusters depend on the large
scale environment.  Loken et al. (\cite{lok99}) have
shown that massive cooling flow clusters are located in high density
regions.  Additionally, X-ray clusters and
clusters with significant substructures (both optical and X-ray
clusters) are located preferentially in high density environments 
(Einasto et al. \cite{e2001}, Plionis \& Basilakos \cite{plb02} and
Schuecker et al. \cite{schu01}). 

Recently the phenomenon of environmental enhancement has  been 
investigated  theoretically using numerical experiments.  Faltenbacher 
et al. (\cite{fal:fal}) and Gottl\"ober et al. (\cite{got:got}) have 
analysed large simulations of the hierarchical formation of 
galaxies, groups and clusters. They studied the average 
mass of galaxy and group size dark matter halos in the 
environment of other galaxies or groups, thereby establishing 
a clear environmental effect extending up to 10 $h^{-1}$Mpc. 
These authors employed the so-called mark correlation functions 
(defined in the cited papers) that generalizes the cumulative 
mass functions employed in the present study. 
The enhanced masses of neighboring halos are connected with
early merging processes giving rise to these halos. Therefore, the
environmental enhancement of the halo masses is a direct evidence for
the process of the hierarchical formation of the galaxy and cluster
halos in a network of filaments that connects high density knots of
the cosmic mass density.

In the present paper we have analysed deep samples of groups of 
galaxies and showed that the properties of loose groups in high 
density environment of rich clusters of galaxies differ from those of 
groups on average --  observational evidence of an environmental 
enhancement of group properties in the gravitational field of 
neighbouring rich clusters.

\begin{acknowledgements} 

We thank Erik Tago and Heinz Andernach for providing us with the 
compilation of the data about Abell clusters, and G\"unther Hasinger 
for providing us with the RBS catalogue. We thank Enn Saar and Sahar 
Allam for stimulating discussions.  The present study was supported by 
Estonian Science Foundation grant 4695 and by Estonian Research and 
Development Council grant TO 0060058S98. P.H. was supported by the 
Finnish Academy of Sciences (grant 46733). J.E. was supported by the 
German Science Foundation (grant 436 EST 17/2/01).  D.L.T. was 
supported by the US Department of Energy under contract No. 
DE-AC02-76CH03000. We thank the anonymous referee for useful comments.

\end{acknowledgements}

\appendix{
\section{ }
\begin{table}
\begin{center}
\caption{The list of LCLG systems around Abell clusters}
\tabcolsep 2pt   
\tiny
\begin{tabular}{rrrrrr}
\hline
$N_{Abell}$ & $\alpha$ &  $\delta$ & $D$ &  $N_{LCLG}$   & Near \\
(1)& (2) & (3) & (4) & (5)&(6)    \\
\hline
\multispan3 Slice $\delta=-3^{\circ}$   &&&\\
\hline
  1098  &  10  45 &  -3  40  &  297.1  &    50, 51    &           \\
  1200  &  11  09 &  -2  53  &  294.6  &    71        &           \\
  1248  &  11  21 &  -3  56  &  319.7  &    85        &           \\ 
  1308  &  11  30 &  -3  42  &  178.2  &    88, 92, 94, 101, 106& \\
  1386  &  11  48 &  -1  40  &  283.7  &   125            &   N   \\
  1399  &  11  48 &  -2  49  &  253.1  &   122, 127, 128  &       \\   
  1404  &  11  49 &  -2  32  &  251.6  &   128, 134       &   N   \\
  1651  &  12  56 &  -3  55  &  236.7  &   172            &       \\
  1658  &  12  58 &  -3  10  &  237.3  &   172, 174, 176  &       \\
  1663  &  13  00 &  -2  15  &  231.9  &   174, 176       &       \\
  1729  &  13  21 &  -3  06  &  316.9  &   185            &       \\ 
  2045  &  15  11 &  -2  34  &  284.0  &   277, 278, 281  &       \\
\hline
\multispan3 Slice $\delta=-6^{\circ}$   &&&\\ 
\hline
   930  &  10  04 &   -5  23  &  235.9  &     4           &     \\      
   978  &  10  18 &   -6  16  &  154.8  &    10, 13, 14   &     \\
  1009  &  10  22 &   -5  32  &  275.9  &    16           &     \\      
  1214  &  11  14 &   -5  20  &  281.4  &    42           &     \\   
  1448  &  12  00 &   -6  33  &  347.9  &    70, 72       &     \\    
  1482  &  12  08 &   -5  18  &  221.5  &    75, 78       &     \\      
\hline
\multispan3 Slice $\delta=-12^{\circ}$   &&&\\ 
\hline
  970   & 10 15  &  -10  27  &   168.6 &     7,  8    & N   \\
 1309   & 11 30  &  -11  34  &   281.6 &    80,  85   &     \\     
 1348   & 11 38  &  -12  05  &   323.5 &    88,  94   &     \\      
 1606   & 12 42  &  -11  43  &   264.9 &   142, 144   &     \\     
 1754   & 13 29  &  -11  24  &   224.2 &   185        &     \\     
 1796   & 13 47  &  -11  40  &   245.2 &   200, 203   &     \\   
 1837   & 13 59  &  -10  55  &   198.1 &   212, 213, 218 &     \\   
 2031   & 15 09  &  -11  00  &  -308.2 &   270, 271   & N   \\
\hline
\multispan3 Slice $\delta=-39^{\circ}$   &&&\\
\hline
 2717   & 00 01  &   -36 14 &  141.7 &   255   & N  \\
 2767   & 00 20  &   -38 24 &  328.3 &   8, 9         &    \\
 2772   & 00 22  &   -38 16 &  327.3 &   9            &    \\ 
 2799   & 00 35  &   -39 24 &  180.9 &  18            &    \\
 2856   & 00 59  &   -38 56 &  336.1 &   28, 29       &    \\
 2860   & 01 01  &   -40 03 &  214.8 &   33           &    \\
 2911   & 01 23  &   -38 14 &  218.8 &   43, 48       &    \\ 
 2960   & 01 57  &   -38 16 &  286.8 &   57           &    \\
 3135   & 03 32  &   -39 10 &  178.2 &  106, 111, 122, 123   &    \\   
 3142   & 03 34  &   -39 58 &  186.7 &   123          &    \\   
 3145   & 03 36  &   -38 11 &  199.2 &   120          &    \\   
 3984   & 23 12  &   -38 04 &  261.5 &   225, 226, 227   &    \\  
 4008   & 23 27  &   -39 36 &  158.1 &   236            &    \\   
 4021   & 23 35  &   -38 23 &  296.1 &   239, 241        &    \\ 
 4029   & 23 41  &   -38 33 &  284.2 &   244             &    \\  
 4068   & 23 57  &   -39 45 &  284.5 &   249             &    \\    
\hline
\multispan3 Slice $\delta=-42^{\circ}$ &&&\\
\hline
 2718   & 00   01  &   -42   13 &  343.7 &     6         &    \\   
 3122   & 03   20  &   -41   31 &  182.6 &   95          &    \\ 
 3756   & 21   12  &   -42   49 &  215.9 &   148, 150    &    \\
 3772   & 21   26  &   -42   58 &  198.4 &   151         &    \\
 3775   & 21   28  &   -43   32 &  291.5 &   156         & N  \\
 3920   & 22   46  &   -41   11 &  345.9 &   212         &    \\  
 3998   & 23   18  &   -42   10 &  252.1 &   236, 237    &    \\   
\hline
\multispan3 Slice $\delta=-45^{\circ}$ &&&\\
\hline
 3102   & 03  12  &   -44  49 &  319.7 &  110        &    \\
 3104   & 03  12  &   -45  36 &  206.7 &  113        &    \\ 
 3111   & 03  16  &   -45  55 &  218.3 &  114        &    \\ 
 3112   & 03  16  &   -44  25 &  211.6 &  114        &    \\ 
 3133   & 03  31  &   -46  07 &  198.4 &  120        &    \\ 
 3749   & 21  06  &   -46  01 &  189.1 &  159        &    \\
 3754   & 21  12  &   -45  41 &  273.8 &  157        &    \\
 3757   & 21  15  &   -45  27 &  267.6 &  167        &    \\
 3809   & 21  43  &   -44  08 &  177.7 &  184        &    \\ 
 3908   & 22  42  &   -45  18 &  252.1 &  217, 219, 223   &   \\
 3910   & 22  43  &   -46  15 &  252.9 &  219, 221        & N  \\
 3963   & 23  01  &   -44  35 &  200.0 &  236, 240        &    \\ 
 3970   & 23  03  &   -45  29 &  345.4 &  239             &    \\
\hline
\label{tab:lga}
\end{tabular}
\end{center}
{\tiny 
{\it Endnotes:}
The columns in Tables~\ref{tab:lga}, \ref{tab:lgapm} 
and \ref{tab:lgx} are as follows:

\noindent Column (1): Number of the cluster from 
Abell catalogue (Table~\ref{tab:lga}), 
from APM catalogue (Table~\ref{tab:lgapm})
and from the RBS catalogue (Table~\ref{tab:lgx}; Rx -- number from the 
catalogue by de Grandi et al. \cite{deg:deg}). 

\noindent Columns (2) and (3): the  right ascension 
(in HH~MM format) and declination (in DD~MM format, equinox 1950)
of rich cluster.

\noindent Column (4): distance of rich cluster, 
in units of ~\Mpc.

\noindent Column (5): LCLG catalogue numbers of loose groups
around clusters. 

\noindent Column (6): Tables~\ref{tab:lga} and \ref{tab:lgapm}: 
index $N$ (for Near) for those clusters that 
have close groups among LCLGs, but which themselves are located 
outside of LCRS slice boundaries. 
Table~\ref{tab:lgx}:  the Abell number of the X-ray cluster
(nA -- non-Abell cluster). 
}                                
\end{table}

}

\end{document}